\pdfoutput=1
\documentclass[autheryear, times]{elsarticle}

\usepackage{pifont} 
\usepackage{self}
\usepackage{amsthm}
\usepackage{booktabs}
\usepackage{times}
\usepackage{url}
\usepackage{float}
\usepackage{algorithm,algorithmic}
\usepackage{algorithm}
\usepackage{tikz}

\usepackage{bm}
\usepackage{mathrsfs}

\usepackage{amsmath}
\usepackage{longtable}
\usepackage{amsfonts}
\usepackage{amssymb}

\usepackage{graphicx}

\usepackage{multirow}
\usepackage{subcaption}
\usepackage{verbatim}
\usepackage{enumitem}
\usepackage{color}
\usepackage{subcaption}

\newlength\myindent
\usepackage{lineno}

\usepackage{bbm} 

\usepackage{hyperref}
\usepackage{xcolor}
\hypersetup{
    colorlinks,
    linkcolor={red!50!black},
    citecolor={blue!50!black},
    urlcolor={blue!80!black}
}

\newtheorem{proposition}{Proposition}
\newtheorem{definition}{Definition}

\usepackage{natbib}

\hypersetup{
    colorlinks,
    linkcolor={red!50!black},
    citecolor={blue!50!black},
    urlcolor={blue!80!black}
}

\makeatletter

\newcommand{\Rmnum}[1]{\expandafter\@slowromancap\romannumeral #1@}
\makeatother

\biboptions{authoryear}

\graphicspath{{images/}}
\journal{}

 \usepackage{setspace}
\begin{document}

\begin{frontmatter}

\title{Estimating Real Demand Using a Flipped Queueing Model: A Case of Shared Micro-Mobility Services}

\author[mymainaddress,address4]{Binyu Yang}
\ead{binyu.yang@connect.polyu.hk}
\author[mymainaddress]{Jinxiao Du}
\ead{jinxiao.du@connect.polyu.hk}
\author[mymainaddress]{Junlin He}
\ead{junlinspeed.he@connect.polyu.hk}
\author[address4]{Shi An}
\ead{anshi@hit.edu.cn}
\author[mymainaddress]{Wei Ma\corref{mycorrespondingauthor}}
\cortext[mycorrespondingauthor]{Corresponding author}
\ead{wei.w.ma@polyu.edu.hk}
\address[mymainaddress]{Department of Civil and Environmental Engineering, The Hong Kong Polytechnic University, Hong Kong SAR, China}
\address[address4]{School of Transportation Science and Engineering, Harbin Institute of Technology, Harbin, China}

\begin{abstract}
The spatial-temporal imbalance between supply and demand in shared micro-mobility services often leads to observed demand being censored, resulting in incomplete records of the underlying real demand.
This phenomenon undermines the reliability of the collected demand data and hampers downstream applications such as demand forecasting, fleet management, and micro-mobility planning.
How to accurately estimate the real demand is challenging and has not been well explored in existing studies.
In view of this, we contribute to real demand estimation for shared micro-mobility services by proposing an analytical method that rigorously derives the real demand under appropriate assumptions. 
Rather than directly modeling the intractable relationship between observed demand and real demand, we propose a novel random variable, Generalized Vehicle Survival Time (GVST), which is observable from trip records. 
The relationship between GVST and real demand is characterized by introducing a flipped queueing model (FQM) that captures the operational dynamics of shared micro-mobility services.
Specifically, the distribution of GVST is derived within the FQM, which allows the real demand estimation problem to be transformed into an inverse queueing problem. 
We analytically derive the real demand in closed form using a one-sided estimation method, and solve the problem by a system of equations in a two-sided estimation method. 
We validate the proposed methods using synthetic data and conduct empirical analyses using real-world datasets from bike-sharing and shared e-scooter systems. 
The experimental results show that both the two-sided and one-sided methods outperform benchmark models. In particular, the one-sided approach provides a closed-form solution that delivers acceptable accuracy, constituting a practical rule of thumb for demand-related analytics and decision-making processes. 

\end{abstract}
\end{frontmatter}

\section{Introduction}
\label{sec:intro}

In recent years, the robust development of the sharing economy has given rise to the emergence of various shared micro-mobility platforms \citep{reck2021uses}. These platforms provide a variety of lightweight vehicles that usually travel at speeds below 45 kph, including shared bikes and scooters \citep{luo2023motivates}. Shared micro-mobility services have gained appeal due to their ability to offer flexible, cost-effective, and on-demand transportation alternatives. They address the need for short-distance travel without relying on private vehicles and also promote sustainable urban transportation \citep{abduljabbar2021role}. 
The increasing recognition of the societal and environmental benefits coupled with advancements in enabling technologies, has led to a rapid growth of shared micro-mobility services worldwide \citep{reck2022mode}.
In the US, over 200 cities have embraced shared micro-mobility services and launched around 150,000 shared fleets into the system \citep{luo2023motivates}. Take the Manhattan area as an example, CitiBike made a significant impact in June 2018 by offsetting over 2.9 million pounds of carbon emissions and contributing to riders collectively burning more than 154 million calories \citep{negahban2019simulation}. 

Despite these advancements, one significant challenge arises from the fact that the data on successful vehicle pick-ups (such as bikes or scooters) in the systems only represents a portion of the overall demand. 
This data typically includes historical trip records and vehicle trajectory data collected by micro-mobility providers.
However, the instances where users were unable to access a targeted vehicle remain unaccounted for. 
In other words, the observable data cannot differentiate between the absence of user arrivals and the presence of a user who arrives but does not take a vehicle. 
This phenomenon indicates that the data may only provide an incomplete view of demand when vehicles are not available in a dock-based station or a service region of dockless systems. 
To clarify this problem, we first define and explain two types of demand for shared micro-mobility services: real demand and observed demand (as detailed in Definition \ref{def:real demand} and Definition \ref{def:observed demand}). 
When the supply side of shared mobility systems (e.g., available vehicles) fails to satisfy the real demand, a portion of the real demand is lost, and the observed demand reflects only the recorded instances of successful vehicle pick-ups. 
That is to say, such observed demand is a censored version of the real demand when there is a loss in real demand due to the lack of vehicles in the service unit within a given period. 
On the other hand, the observed demand equals real demand when the supply side consistently meets real demand. 
However, vehicle unavailability is common in practice, particularly during peak hours, when the supply is insufficient to meet the high demand, thereby preventing some users from accessing a vehicle. 
Consequently, the available supply in the service unit serves as an upper limit for the observed demand \citep{huttel2023mind}.

\begin{definition}[Real demand]
\label{def:real demand}
Consider a service unit, which could either be a station in a dock-based system or a service region in a dockless system. 
The real demand refers to the number of users arriving to pick up vehicles during a specified time period, assuming there are no supply constraints.
\end{definition}

\begin{definition}[Observed demand]
\label{def:observed demand}
The number of successful pick-ups made by users as recorded in the operational data of shared mobility services is referred to as observed demand.

\end{definition}

\begin{figure}[h]
    \centering
    \includegraphics[width=1\linewidth]{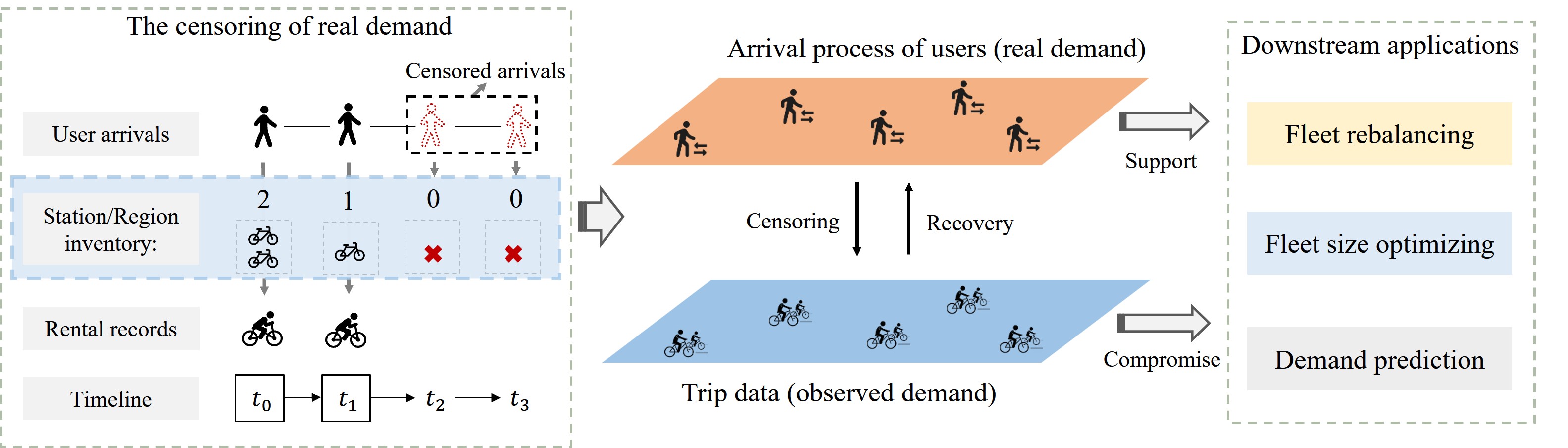}
    \caption{Illustration of the real demand in the micro-mobility system.}
    \label{fig:intro}
\end{figure}

Although the issue of observed demand being censored is acknowledged, there are still many studies that ignore the differences between the observed demand and real demand. 
In current demand-driven studies, the observed demand from data serves as an approximation of real demand due to limited availability and challenges in collecting real demand data \citep{schuijbroek2017inventory, li2022mean}. 
However, such underestimated demand can lead to suboptimal decision-making in various scenarios. 
In particular, machine learning models that rely solely on observed demand data from shared micro-mobility services consider only a censored version of demand, resulting in biased predictions. 
This bias leads to the underestimation of real demand, thereby restricting the ability of operators to effectively plan operations and manage the system \citep{gammelli2020estimating,huttel2022modeling, huttel2023mind}. 
Conversely, accurately uncovering real demand could provide a solid foundation for addressing decision-making challenges, including optimizing fleet size \citep{luo2020optimizing}, developing effective rebalancing strategies \citep{legros2019dynamic}, and demand prediction \citep{lin2018predicting}.
The left module in Figure \ref{fig:intro} illustrates how a portion of the real demand is censored and lost in observed records when vehicles are unavailable in a service unit. 
The middle module illustrates that the latent user arrival process (including both the lost and observed users for vehicle pick-ups) is covered by the observed trip data from operators due to the lack of available vehicles, and the observed demand becomes a censored version of the real demand. 
The right module indicates that neglecting real demand can lead to ineffective policies and strategies.
In light of the above problems caused by directly utilizing the observed demand to replace the real demand, it is crucial for both public agencies (e.g., Transport Department) and private entities (e.g., Shared Micro-mobility Companies) to accurately recover the real demand to inform effective management and operational strategies for shared micro-mobility services. 

Estimating the real demand for shared micro-mobility services presents significant challenges.
The main reason is that directly modeling the difference between the observed demand and real demand is intractable, as the information directly pertaining to the lost portion of demand remains inaccessible. 
Some studies regarded observed demand as censored observations of the real demand, employing censored modeling techniques (e.g., the integration of machine learning models with the Tobit model) to estimate the real demand \citep{gammelli2020estimating, huttel2022modeling}. 
However, these data-driven methods rely on the capabilities of machine learning-based models, which are theoretically intractable and are unable to explain the underlying mechanisms that generate the observed demand. 
In addition, the performance of censored modeling approaches may be diminished when supply remains insufficient over an extended period, resulting in the majority of observations being censored. 
Another line of studies seeks to recover the real demand by leveraging external information, including Point-of-interests (POIs) and migration behaviours \citep{wang2023recovering, wang2023demand}. 
Nonetheless, numerous external factors contribute to the generation of real demand, which complicates the modeling and verification of complex interrelationships between these factors and demand in practical applications \citep{eren2020review}. 
In summary, existing studies focus on directly modeling the relationship between the observed demand and real demand, which either rely on intractable models or necessitate additional assumptions regarding the real demand generation process.
This limitation indicates that the recovery of real demand may provide only a partial representation, rather than a complete characterization of real demand.

To address the above gaps, we propose an analytical method to estimate the real demand, which is theoretically tractable and supports hypothesis testing.
Notably, this approach neither relies on external data nor assumes that the generation of real demand is influenced by external factors. 
We only assume the Poisson arrival process of the users and vehicles, which are widely adopted in the literature \citep{raviv2013optimal, schuijbroek2017inventory, wang2023recursive}. 
To this end, we propose a novel observable random variable, named Generalized Vehicle Survival Time (GVST) as a proxy for the observed demand, and construct the relationship between the real demand and the GVST by developing a tractable flipped queueing model (FQM).
In this manner, we are able to analytically derive the real demand rather than using optimization-based or machine learning-based methods, which is different from all state-of-the-art studies.
Specifically, the FQM is developed to capture the dynamics of micro-mobility services, where the demand side of micro-mobility services is flipped to be modeled as the supply side of a queueing process. 
By applying the FQM, the distribution of GVST can be derived explicitly, which transforms the real demand estimation problem into an inverse queueing problem \citep{asanjarani2021survey}. 
Specifically, we introduce both two-sided and one-sided estimation method to solve this problem, enabling the analytical recovery of the real demand based on GVST observations.
The two-sided method derives the real demand by solving the system of equations, while the one-sided method analytically calculates the real demand in closed form.
Given the unobservable nature of the real demand, validating the proposed methods poses challenges. 
Therefore, we verify our methods by applying them to synthetic datasets generated through simulation experiments \citep{negahban2019simulation} and analyzing the estimated results using real-world datasets from bike-sharing and e-scooter systems. 
The results demonstrate that our estimation methods outperform state-of-the-art models and provide reasonable estimations in real-world scenarios. 
In addition, our proposed method can be applied to both dock-based systems where users are required to pick up vehicles at designated stations, and dockless systems where vehicles can be accessed at various locations.
The contributions of this paper are threefold:
\begin{itemize}
    \item This research highlights the task of real demand estimation for shared micro-mobility services. 
    By novelly developing the concept of generalized vehicle survival time (GVST), we transform the real demand estimation task into an inverse queueing problem.
    \item 
    A flipped queueing model is proposed to model the dynamics of shared micro-mobility services and derive the distribution of GVST. 
    Without applying machine learning or optimization-based methods, the real demand can be analytically derived by solving a system of equations in the two-sided method, while the one-sided method provides a closed-form estimation of the real demand. 
    \item The proposed methods are verified through synthetic datasets generated from simulation experiments. Real-world datasets from bike-sharing and e-scooter companies are applied to test the performance. 
    The results demonstrate the effectiveness of our methods in accurately recovering the real demand.
    The closed-form solution can be treated as a reliable rule of thumb for people to approximate the real demand for shared micro-mobility services without complex numerical calculations.
\end{itemize}

The remainder of this paper is organized as follows: related literature is reviewed in Section \ref{sec:related}. Section \ref{sec:model} presents the formulations of the proposed method to estimate the real demand, and Section \ref{sec:Experiments} shows the experimental results on synthetic datasets and the real-world applications. Finally, the conclusion is drawn in section \ref{sec:Conclusion}.

\section{Related work}
\label{sec:related}
In this section, we first introduce the studies on real demand estimation for shared economies and then review the queueing models applied in shared micro-mobility services and the studies for the inverse queueing problem.

\subsection{Real demand estimation}

The problem of estimating the real demand for shared mobility services has drawn increasing attention in recent years.
For ride-hailing systems, \citet{fields2021data} simulated the car-sharing demand truncation process with different levels of demand distribution, then compared the output of the simulation and the pick-ups in historical data to determine the most likely real demand distribution. 
The demand migration behavior was implicitly considered in the simulation rather than explicitly formulated in the analytical model.
Using taxi GPS data in New York City, \citet{zhang2019demand} applied queueing theory to derive an analytical equation that relates the observed demand, taxi demand, customer waiting time, and taxi supply. They assume the arrivals of both customers and vacant taxis follow Poisson distributions, customer waiting times follow an exponential distribution, and taxis pick up customers based on the First In First Out rule. The demand for street-hail taxis in each road segment is estimated using the non-stationary Poisson random field method.
Similarly, by utilizing street-hail taxi data in Xian City, a recursive decomposition probability model is developed to estimate the real demand by incorporating the disaggregated passing time of each taxi \citep{wang2023recursive}.

For micro-mobility services, existing studies have been conducted on both dock-based systems and dockless systems. \citet{albinski2018performance} calculated the mean value of demand that is not censored in the same period to replace observed demands to evaluate the service levels of dock-based bike-sharing and dockless bike-sharing systems. 
In addition, based on the CitiBike dataset, a discrete-event simulator is developed to estimate the demand between each pair of stations in New York City \citep{jian2016simulation}.
Similarly, an iterative simulation-based inference method was established by Negahban to estimate the real demand for dock-based bike-sharing, proving that ignoring the demand censoring causes incorrect or suboptimal decisions \citep{negahban2019simulation}. To estimate the demand for dockless bike-sharing, \citet{wang2019locally} treated censored observed demand as the missing value in the matrix and established a matrix completion solution to recover these values. 
Focusing on the dockless bike-sharing, \citet{wang2023demand} and \citet{wang2023recovering} both assume the observed and real demand follow Poisson distributions, and the unmet demand will migrate to the nearby areas. The real demand of each area is assumed to be related to the contained POIs, and the demand function was formulated based on the observed demand \citep{wang2023demand}. The MLE method is applied to estimate the impact of changes in Points of Interest (POIs) on demand and to what extent the changes in influential factors (i.e., metro station number) influence user migration to adjacent grids.
In the follow-up work, the real demand recovery is modeled as a maximum-a-posteriori (MAP) problem \citep{wang2023recovering}. Specifically, the prior term is computed by defining the spatial correlation function between a target area and the nearby eight areas. The conditional term models the distribution between the observed demand and expected real demand, considering the demand truncation and migration.
Subsequently, a tailored simulated annealing approach is employed to solve the optimization problem defined by the MAP framework.

Moreover, the Tobit model is a common method applied for censored cases, and a representative study on bike-sharing was conducted by \citet{gammelli2020estimating}, where the authors integrated the Tobit model with a Gaussian process and modeled the problem of real demand recovery as a time-series analysis to recover the real SBBS demand. The following work makes no assumptions on the parametric form of the latent demand distribution by applying Multi-Output Censored Quantile Regression Neural Networks \citep{huttel2022modeling}. In this way, multiple quantiles of the predictive distribution are estimated while accounting for censorship in the observed demand. \citet{huttel2023mind} further extends these censorship-aware models to both the temporal and spatial dimensions by employing temporal graph neural networks. 
The proposed model considers the accessibility of adjacent charging facilities and temporal features when estimating the real electric vehicle charging demand.

However, the methods for ride-hailing are not transferable to the micro-mobility services due to the disparate data formats.
Some studies require that the input data include an indicator denoting whether the demand is censored or not \citep{gammelli2020estimating, huttel2022modeling}. This leads to the neglect of analyzing the information contained in the observed arrival process of users and vehicles.
Other studies assume that apart from the real demand, the observed demand also follows a Poisson distribution, which may be impractical, especially in scenarios where demand significantly exceeds supply and the real demand is censored heavily. 
In addition, extra information is required in these methods to estimate the real demand such as POIs \citep{wang2023demand} or the preset parameter related to the migration behaviours and spatial correlation extent \citep{wang2023recovering}. 
Our proposed method avoids extra assumptions about the distribution of observed demand and external effects on real demand, which provides an analytic estimation of real demand without extra data.

\subsection{Queueing models for shared micro-mobilities}

Fundamental notations of queueing theory are first presented for clarity. A queuing system is typically denoted as M/D/1/K when the following assumptions are satisfied. The arrival of customers follows a Poisson process. 
The service time of customers is fixed and constant, denoted by $D$;
The system only serves one customer (i.e., one server) at one time with a fixed deterministic time; The system has a finite capacity, denoted by $K$. The traffic intensity $\rho$ as a measure of congestion is computed by dividing the customer arrival rate by the product of the service rate and the number of servers.

We now discuss the application of queueing models in micro-mobility systems. Queueing theory and Markov processes have been applied to model the bike-sharing system \citep{fricker2012mean, leurent2012modelling}. Most of these studies focus on the system performance measures \citep{leurent2012modelling, ekwedike2021group} and rebalancing strategies \citep{schuijbroek2017inventory} of the bike-sharing system. In this context, the M/M/1/K queue is usually applied to model the bike-sharing system \citep{fricker2012mean, fricker2016incentives, schuijbroek2017inventory, ekwedike2021group, tao2021stochastic}. The arrivals of users and vehicles are assumed to follow the Poisson process so that the inventory of stations with capacity can be modeled as a Markov chain. In this way, \citet{raviv2013optimal} estimates the transient distribution of M/M/1/K and obtains the expected number of bike shortages at the station level. \citep{schuijbroek2017inventory} evaluate the service level of each station by modeling them as M/M/1/K queues and generate the strategies for inventory rebalancing.
\citet{ekwedike2021group} derive the exact solutions for transition probabilities of the M/M/1/K queue using group symmetry methods and apply them to the bike-sharing context.

Instead of studying the bike-sharing system at a station level, some studies apply closed queueing networks to model the whole system and focus on macro system performances \citep{tao2021stochastic}. 
One of the seminal papers to model the whole system as a closed queueing network is \citet{george2011fleet}. The queueing network includes typical station nodes modeled by M/M/1/K and virtual nodes modeled by M/M/1. Each virtual node connects two station nodes and the service times of virtual nodes represent the trip durations from one station to another. In the follow-up work, \citet{li2016unified} provide a unified closed queueing network, in which the stations and roads between stations are simplified into indistinct nodes (A routing matrix is constructed to compute the arrival rate in each node).
Similarly, \citet{li2017nonlinear} consider a more complicated queueing network where the arrival pattern and spatial structure of actual systems are included. They focus on obtaining the stationary probability of problematic (fully occupied or empty) stations based on the stationary distribution of joint queue lengths. 
The failure, removal, repair, and redistributing processes of bikes are included in the queueing network proposed by \citet{fan2022dockless} using RG-factorizations of block-structured Markov processes. 
\citet{benjaafar2022dimensioning} extend the closed queueing network to dockless systems and provide closed-form bounds on the optimal vehicle number under various service levels. 

To solve the vehicle routing problem, \citet{schuijbroek2017inventory} assess the quality of bike-sharing service by using an M/M/1/K queue and establish optimization models for static rebalancing. 
The flow equivalent server algorithm is employed in \citet{fan2020optimization} to ease the computation by partitioning the queueing network into disjoint subnetworks and determining the repositioning flow. \citet{mo2021modeling} also contribute to the rebalancing strategy by distinguishing higher-demand and lower-demand nodes. 
Compared to the research on bike-sharing systems, the body of literature concerning electric scooters remains underdeveloped. \citet{pender2020stochastic} model the dynamics of battery life within a scooter network using queueing models. \citet{liu2024planning} introduce a queuing network model to model the charging and rider engagement process for shared dockless micro-mobility services.

The above-mentioned studies all treat the arrival rates of users and vehicles as the input of models. These arrival rates are usually determined by simulation \citep{fricker2016incentives, mo2021modeling, fan2022dockless, liu2024planning}, computed using mean field theory \citep{tao2021stochastic}, or approximated from observed demand data \citep{schuijbroek2017inventory}, or both \citep{raviv2013optimal, li2022mean}. 
However, such arrival rates do not represent the real demand as discussed in Figure \ref{fig:intro}. Differently, this paper flips the demand side and supply side by applying a flipped queue model to view the observable variables as input and concentrates on the estimation of the real arrival process.
The proposed method is validated through simulations, consistent with the approach used in the above studies.

\subsection{Inverse queueing problem}

The inverse problem of queueing systems is to estimate or predict the exogenous processes (i.e., customer arrival and service process) based on the observed endogenous processes (e.g., queue length, waiting time, or departure process) \citep{asanjarani2021survey}.
In this scenario, specific attributes of the system are monitored, and the resultant observations are subsequently employed to conduct parameter inference. 
\citet{basawa1996maximum} use maximum likelihood estimation (MLE) to estimate the arrival and service rates based on observed waiting times. The performance on M/M/1 and M/$E_k$/1 queues shows that the results are biased with less than 5000 samples of waiting time. In follow-up work \citep{ross2007estimation}, the MLE method is extended to the case of multiple servers based on observed queue length data. The maximum likelihood estimators of mean arrival rate and different service rates for heterogeneous servers are derived in \citet{wang2006maximum}. \citet{hansen2006nonparametric} solve the inference problem of M/G/1 systems using sampled observed workloads, which is the aggregate of the remaining service durations for the customer being served and those in the queue awaiting. Further studies extend the estimation of service time distribution to infinite server queues. The observed arrival–departure data \citep{goldenshluger2018m} and queue-length process \citep{goldenshluger2016nonparametric} are applied to estimate the service time distribution using nonparametric methods. Based on the same observations, \citet{goldenshluger2019nonparametric} relax the assumption of the arrival process to a nonhomogeneous Poisson process.

In the cases when the endogenous processes are not observed directly, the empirical data can be obtained by conducting system probing through artificial customers. Based on the measurements of probing customers (e.g., sojourn times), the solution to the inverse problem can be obtained. One of the seminal works focuses on the M/D/1 system by assuming an independent arrival stream is known \citep{chen1994parameter}. Maximum likelihood estimation of the arrival rate is achieved using the observed arrival and waiting times.
\citet{alouf2001inferring} extend the study to the systems with limited capacity $K$, M/D/1/K and M/M/1/K. Eleven moment-based estimators of the probe stream are applied to infer the arrival rate and $K$. 
A lightweight probing method for estimating arrival rates based on observed interarrival and inter-departure times is proposed in \citet{hei2005light}, and different efficient characteristics of the probing stream are also discussed for changing intensity $\rho$. Subsequently, various probing methods are developed for the estimation of available bandwidth (the residual processing capacity) \citep{hei2006model, kauffmann2012inverse} and parameters of queueing networks \citep{baccelli2009inverse, incerto2018moving}. Recently, real-world data have been employed to recover the arrival process. A simulation-based method using appointment and observed arrival data is applied to model the arrival process to a clinic \citep{kim2018data}. More details of these methods can be found in the survey conducted by \citet{asanjarani2021survey}.

In summary, this paper aims to fill the current research gaps and estimate the real demand for micro-mobility services.
We propose an FQM to represent the micro-mobility system and formulate the real demand estimation problem as an inverse queueing problem. 
Solving this problem refers to estimating the exogenous arrival processes of users and vehicles in our case, by observing the endogenous process of the proposed FQM, which is the GVST in our study. 

\section{Model}
\label{sec:model}

In this section, we first clarify the real demand estimation task for shared micro-mobility services. Then, a new random variable GVST is introduced. The FQM is proposed to model the distribution of this variable. 
Both two-sided and one-sided methods are applied to estimate the real demand based on the observations of GVST. Lastly, the computational steps are presented.

\subsection{Settings}

Following the former literature \citep{raviv2013optimal, schuijbroek2017inventory}, user arrivals for vehicle pick-ups and vehicle arrivals for drop-offs are assumed to follow two independent Poisson processes. 
This assumption aligns with findings from \citet{o2015data} and \citet{negahban2019simulation} on micro-mobility data. 
During a specific period (e.g., one hour), we assume the arrival rates for vehicle pick-up and drop-off in a station or region are constant and homogeneous within the time period \citep{wang2023demand}. 
To enhance clarity, we refer to both a station and a region as a `unit', which constitutes the fundamental units of estimation and analysis in our method. 
That is to say, the real demand estimation is conducted at the unit level.
The real demand as a random variable is denoted by $N_r$, representing the number of users arriving at a unit for vehicle pick-ups during the period without supply constraints.
According to the Poisson process assumption, the real demand $N_r$ follows a Poisson distribution with rate $\mu$, where $\mu$ represents the arrival rate of the users. 
This study focuses on estimating the rate parameter $\mu$ of this underlying Poisson distribution, namely the expected value of the real demand, based on the observed data.

The observed data is usually the trip records provided by shared micro-mobility providers. 
A typical data structure of the trip or order records contains the vehicle Identification Number (ID), the times when the users picked up the vehicles and the pick-up locations, the times when the users dropped off the vehicles and the drop-off locations. 
Given a unit and a time period, we can filter the data to include trips where either (1) the pick-up location corresponds to the target unit and the pick-up time falls within the period, or (2) the drop-off location corresponds to the target unit with the drop-off time within the period.
Then, a dataset of vehicle IDs along with the corresponding times when they were dropped off at and/or picked up from the unit can be obtained. 
In another commonly used data structure, each row corresponds to a specific vehicle ID, the observed timestamp, the associated location information (usually represented by latitude and longitude), and the status indicating whether the vehicle is idle or in use.
By defining the transition from `idle' to `in use', and from `in use' to `idle', the pick-up and drop-off events of each vehicle can be captured. In this way, the data structure can be transformed into the format of trip data and then obtain the dataset.
This dataset serves as the input of our proposed method, which is denoted by $\mathcal{D} = \{(I_n, t^*_n, k_n)\}_{n=1}^N$, where $I_n$ is the vehicle ID of $n^{\mathrm{th}}$ record, $t^*_n$ corresponds to either the pick-up or drop-off time depending on the value of $k_n$, and $k_n$ is a binary indicator, with 0 indicating a pick-up event and 1 indicating a drop-off event, $N$ is the total number of observed pick-ups and drop-offs.

\begin{figure}[h]
    \centering    \includegraphics[width=0.9\linewidth]{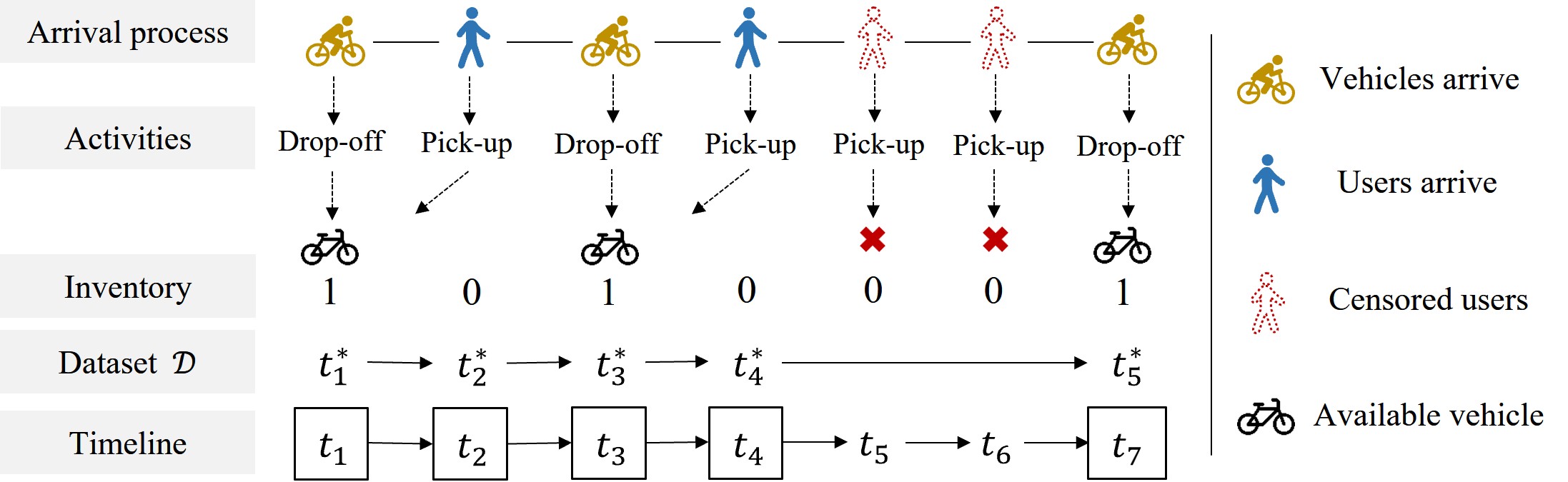}
    \caption{An example for the arrival processes.}
    \label{fig:Arrival}
\end{figure}

The limitation of observed data $\mathcal{D}$ in accurately capturing the real demand is illustrated in Figure \ref{fig:Arrival}, which presents an example in which users and vehicles arrive at a unit.
It can be seen that the user arrives at time $t_5$ and $t_6$ seeking vehicle pick-up are aborted due to insufficient vehicle availability in the inventory.
As a result, the demand recorded in the observed data $\mathcal{D}$ underestimates the real demand. 
To this end, this study aims to solve this problem and recover the real demand based on the observed data.
Specifically, taking the observed data $\mathcal{D} = \{(I_n, t^*_n, k_n)\}_{n=1}^N$ as input, our method aims to estimate the expected value of the real demand, denoted by $\mu$.

\subsection{Generalized vehicle survival time (GVST)}
\label{sec: GVST}

Instead of building the direct connection between the observed demand and real demand, we introduce a new random variable named generalized vehicle survival time (GVST), denoted by $Y$. 
It captures latent information related to the real demand, and the observations of GVST can be computed from the observed data $\mathcal{D}$. 
Firstly, we discuss vehicle survival time, which refers to the duration a vehicle remains within a unit.
This vehicle survival time can be calculated from $\mathcal{D}$ by subtracting its drop-off time from its pick-up time for each individual vehicle.
Let $\boldsymbol{t_d}$ denote the vector of drop-off times corresponding to each vehicle in the drop-off sequence, where $t^1_d \leq t^2_d \leq \cdots \leq t^{N_d}_d$ and $N_d$ is the total number of observed drop-offs.
For the $i^{\mathrm{th}}$ dropped off vehicle, the corresponding pick-up time is denoted by $t^i_p$, and the vehicle survival time is computed as $t^i_p - t^i_d$. 
However, the modeling of vehicle survival time is intractable since the arriving users randomly pick up the available vehicles. 
To this end, we propose the Generalized vehicle survival time as defined in Definition \ref{def:gvst}.

\begin{definition}[GVST]
\label{def:gvst}
The GVST is defined as the elapsed time between the moment a vehicle is dropped off and its subsequent pick-up time following the implementation of a swap scheme. The GVST of $i^{\mathrm{th}}$ vehicle is computed as:
\begin{align} 
\label{eq:observed drop-offs}
\boldsymbol{t_d} &= (t^1_d, \cdots, t^{N_d}_d) = sorted\left\{ t^*_n \;\middle|\; (I_n, t^*_n, k_n) \in \mathcal{D},\, k_n = 1 \right\}, \\
\label{eq:observed pick-ups}
\boldsymbol{\tilde{t}_p} &= -th(\tilde{t}^1_p, \cdots, \tilde{t}^{N_p}_p) = sorted\left\{ t^*_n \;\middle|\; (I_n, t^*_n, k_n) \in \mathcal{D},\, k_n = 0 \right\}, \\
\label{eq:gvst}
y_i &= \tilde{t}^{h_i}_p - t^{i}_d,\;\; h_i =\min \left\{ j \;\middle|\; \tilde{t}_p^j > t^{i}_d,\, j = h_{i-1} + 1, h_{i-1} + 2, \ldots, N_p \right\},
\end{align} 
where $\mathrm{sorted}\left\{\cdot\right\}$ denotes sorting the elements in chronological order; 
$\boldsymbol{t_d}$ denote the vector of drop-off times corresponding to each vehicle in the drop-off sequence;
The vector $\boldsymbol{\tilde{t}_p}$ represents the sequence of observed pick-up times, such that $\tilde{t}^1_p \leq \tilde{t}^2_p \leq \cdots \leq \tilde{t}^{N_p}_p$, and $N_p$ denotes the total number of observed pick-ups.
$y_i$ denotes the GVST for the $i^{\mathrm{th}}$ vehicle; $\tilde{t}^{h_i}_p$ represents the pick-up time of $i^{\mathrm{th}}$ vehicle after applying the swap scheme, which indicates the time of the first user arrives for pick-up after time $t^{i}_d$; $h_0$ is set as 0.
\end{definition}

\begin{figure}[h]
    \centering    \includegraphics[width=0.8\linewidth]{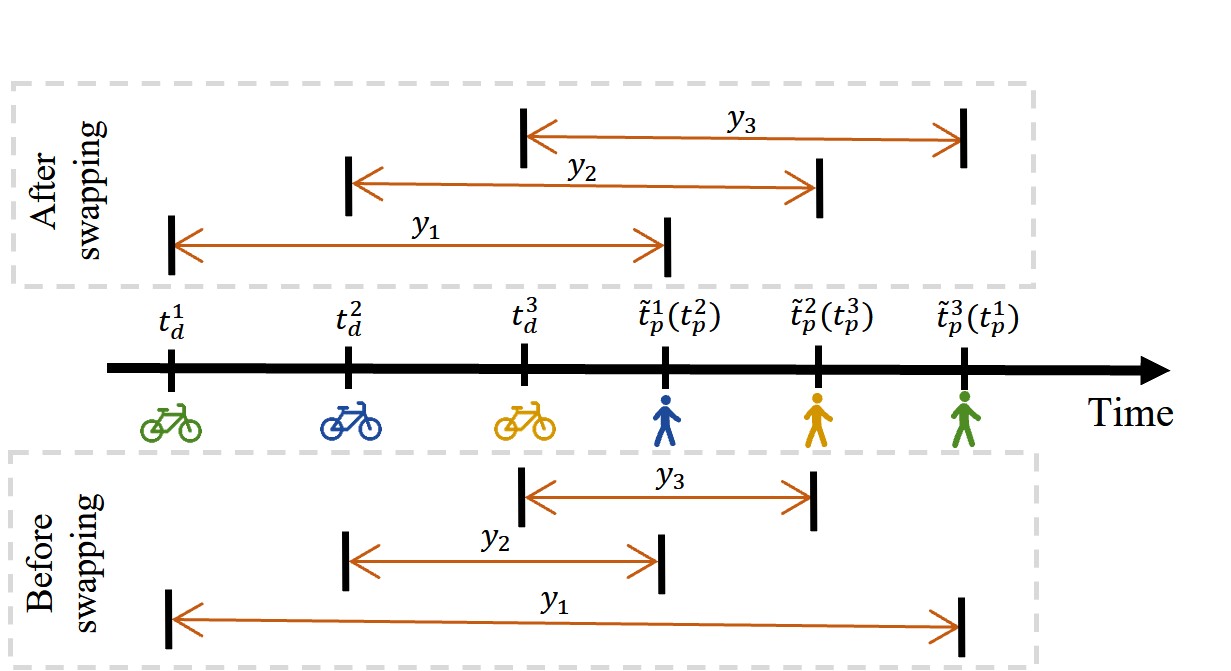}
    \caption{The collection process of GVST observations.}
    \label{fig:GVST}
\end{figure}

Specifically, the swap scheme is applied to the observed pick-up time of vehicles, ensuring that the collection of GVST observations follows the criteria that each arriving user always picks up the vehicle that was dropped off earliest. 
An example of such a scheme is shown in Figure \ref{fig:GVST}, where $t_d^i$ is the drop-off time of $i^{\mathrm{th}}$ vehicle, $t_p^i$ is the observed pick-up time of $i^{\mathrm{th}}$ vehicle, and $\tilde{t}_p^j$ is the arrival time of $j^{\mathrm{th}}$ user for pick-ups.
According to the definition of GVST, the pick-up time of the first vehicle is swapped from $t_p^1$ to $\tilde{t}_p^1$ since $\tilde{t}_p^1$ is the arrival time of the first user who arrived after $t_d^1$.
For the second GVST observation, the pick-up time of the second vehicle is swapped to the arrival time of the first user after $t_d^2$ excluding the users who have already picked up vehicles, namely $\tilde{t}_p^2$.
The collection of GVST observations concludes when all vehicle pick-up times have been swapped or there are no new users arriving subsequent to the arrival time of a vehicle during the period.

The collected GVST observations are denoted by $\boldsymbol{y}=\{y_i\}^{N_g}_{i=1}$, where $N_g$ is the total number of GVST observations. 
Note that the GVST is observable for any period where at least one pick-up after a drop-off happens and therefore can be easily collected from trip data that records user and vehicle arrival times of the unit.
To model the GVST within the shared micro-mobility systems, we propose a flipped queueing model, which will be elaborated in the next section.

\subsection{The flipped queueing model (FQM)}
\label{sec: Flip}

In this section, we introduce the proposed flipped queueing model (FQM) characterizing the operational dynamics of shared micro-mobility services and build the relationship between GVST and FQM. 
In FQM, the users arriving for pick-ups as the demand side of the micro-mobility service is flipped to be modeled as the supply side of a queueing process. 
Accordingly, the arriving vehicles as the traditional supply side, are reinterpreted as the demand side of the queue. 
Recall that both user arrivals for pick-ups and vehicle arrivals for drop-offs are assumed to follow independent Poisson processes, and the arrival rate of users is $\mu$. Let $\lambda$ denote the arrival rate of vehicles for drop-offs.
In the FQM, the unit is constrained by a finite available vehicle capacity, denoted by $K$. 
Vehicles arriving at the unit will either wait in the queue for pick-ups or be removed from consideration if the number of available vehicles is less than $K$; Conversely, they will abort and leave the unit once the number of available vehicles reaches $K$. 
Arriving users are considered to provide `services' for the vehicles. 
If there are vehicles available in the unit, the user will pick up a vehicle and the inventory (number of available vehicles) will decrease by one. 
Otherwise, the users are aborted immediately and therefore lose the demand. 
The time for vehicle pick-up and drop-off is assumed to be zero, namely negligible. 

Through this formulation, the FQM effectively captures the dynamics between the demand and supply sides of the shared micro-mobility services. 
We further formulate the Markov chain for the FQM during the period. 
Let $X$ be the random variable representing the inventory of vehicles at a given unit. 
The stochastic process $\{X(t): t \geq 0\}$ describes the evolution of the vehicle inventory over time, where $X(t)$ denotes the number of vehicles present at time $t$. We model this process as a continuous-time Markov chain, with vehicle drop-offs occurring at rate $\lambda$ (the birth rate) and vehicle pick-ups occurring at rate $\mu$ (the death rate). Furthermore, the state space of this Markov chain is finite, with the maximum inventory level $K$. 
According to the Definition \ref{def:gvst} of GVST, GVST can be viewed as an output variable of the proposed FQM:
\begin{equation}
    Y \sim FQM(\mu, \lambda),
\end{equation}
where $Y$ is the GVST; $FQM(\lambda, \mu)$ denote the FQM with vehicle arrival rate $\lambda$ (supply side) and user arrival rate $\mu$ (demand side) as its input parameters.

Then, the FQM can be modeled as a flipped M/M/1/K queue, characterized as:
\begin{equation}
  \vcenter{
    \hbox{
      \begin{tikzpicture}
        \node (eq) at (0,0) {$FQM(\lambda, \mu) = M/M/1/K(\mu, \lambda)$,};
        \draw[<->] ([xshift=1.2cm, yshift=-0.1em] eq.south west)  
          -- ++(0,-0.25cm) 
          -- ++(2.8cm,0) 
          -- ++(0,0.25cm)
          node[midway,below,xshift=-1.4cm, yshift=-0.2cm, align=center, font=\scriptsize]
          {Flip the demand and supply side};
      \end{tikzpicture}
    }
  }
\end{equation}
where $M/M/1/K(\mu, \lambda)$ represents an M/M/1/K queueing system in which the typical roles of demand and supply sides are flipped: the service rate is determined by the demand rate $\mu$ and the customer arrival rate is determined by the supply rate $\lambda$. 

\begin{figure}[ht]
    \centering    \includegraphics[width=0.9\linewidth]{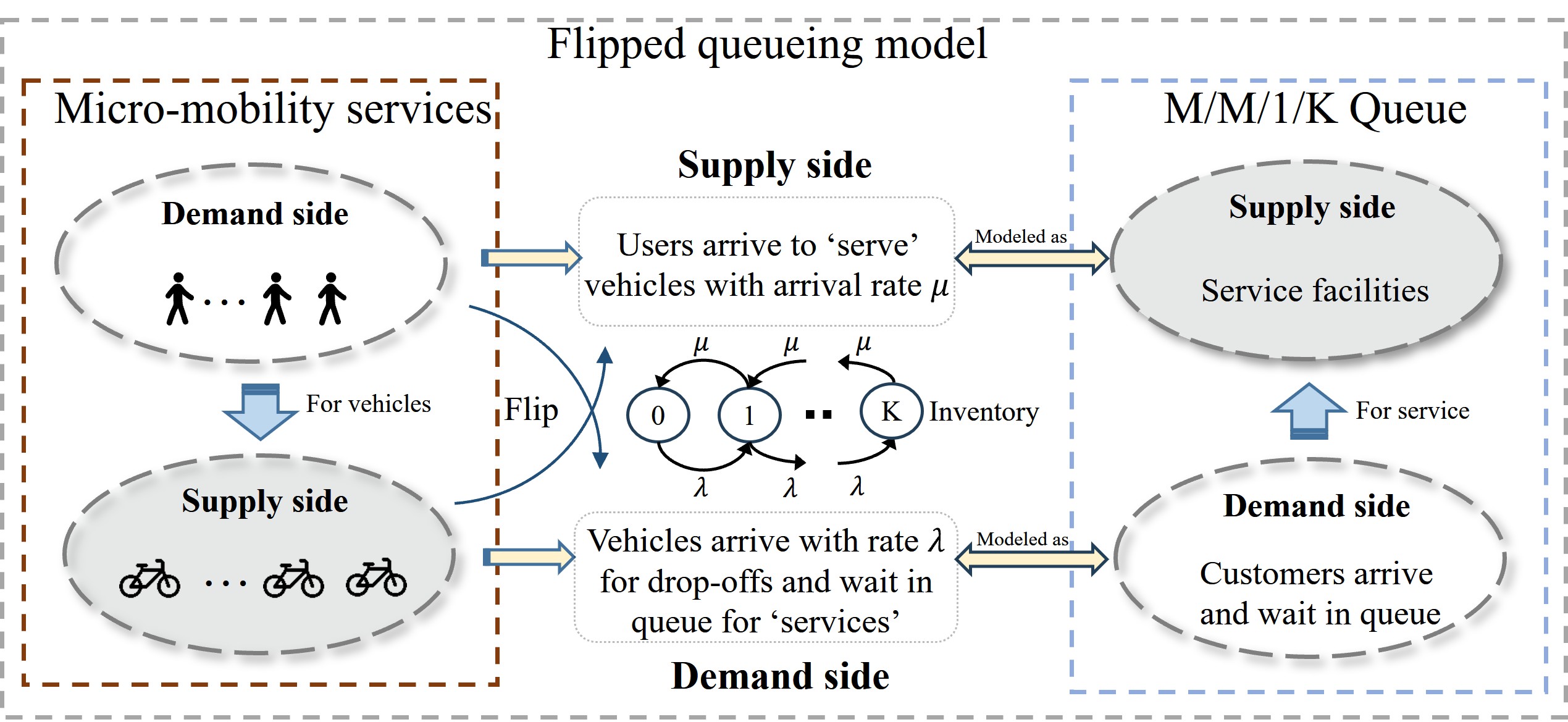}
    \caption{The flipping of demand and supply side in the FQM.}
    \label{fig:flipped model}
\end{figure}

Specifically, the relationship between the FQM and M/M/1/K queue is shown in Figure \ref{fig:flipped model}. 
One can see that the demand side in shared micro-mobility services (e.g., arriving users for picking up vehicles in FQM) is modeled as the supply side of M/M/1/K queue (e.g., users arrive to provide `services' for the vehicles that are waiting in the queue). 
Here, the `service' indicates the pick-up action. 
On the other hand, the supply side (e.g., arriving vehicles for drop-offs) of FQM is modeled as the demand side of the M/M/1/K (e.g., the vehicles being dropped off will wait in the queue for `services'). 
To summarize, the FQM can be modeled as an M/M/1/K queue where the `service' time (inter-arrival time of users) is exponentially distributed with mean $1/\mu$, and the arrival rate of `customers' (vehicles) is $\lambda$. 
To distinguish between the FQM and the M/M/1/K, the `customers' mentioned in this paper only denote the ones seeking services in the M/M/1/K queueing model, while the ones who arrive aiming to pick up vehicles in our model are denoted as `users'.
Table \ref{tab:notations} summarizes the notations used in this paper.

The proposed GVST as the output of the $FQM(\mu, \lambda)$ can be viewed as the observable endogenous processes of the queueing model and our objective is to estimate the exogenous process (user arrival process).
Then the real demand estimation problem can be transformed into the inverse queueing problem.
To this end, we aim to build the relationship between GVST and user arrival rate $\mu$ by reaching the distribution of GVST within the FQM, which will be explained in the next section.

\begin{table}[ht]
   \centering
   \renewcommand{\arraystretch}{1.25}
    \footnotesize
   \caption{Summary of notations}
   \begin{tabular}{cll}
   \toprule
   \textbf{Notation} & \textbf{Micro-mobilities} & \textbf{M/M/1/K queue} \\
   $\lambda$ & Arrival rate of vehicles for drop-offs & Arrival rate of customers for service \\
   $\mu$ & Arrival rate of users for vehicle pick-up & Expected service rate \\
   $K$ & The vehicle capacity of the unit &  Queue capacity including the one in service \\
   $X$ & Number of available vehicles (inventory) of the unit & Number of customers in the system \\
   $P_x$ & Probability that exactly $x$ vehicles are available & Probability that $x$ customers in the system \\
   $Y$ & GVST & Sojourn time of a customer in the system \\
   $F_Y(y)$ & CDF of GVST & CDF of waiting time in the system \\
   $f_Y(y)$ & PDF of GVST & PDF of waiting time in the system \\
   $\rho$ & Traffic intensity & Traffic intensity \\
   $\boldsymbol{\tilde{t}_p}$ & Observed time epochs that users arrive for pick-ups & \multicolumn{1}{c}{$\slash$} \\
   $\boldsymbol{t_d}$ & Observed time epochs that vehicles arrive for drop-offs & \multicolumn{1}{c}{$\slash$} \\
   $\boldsymbol{y}$ & Observations of GVST & \multicolumn{1}{c}{$\slash$} \\
   $N_r$ & The real demand & \multicolumn{1}{c}{$\slash$} \\
   $N_p$ & Number of observed arrival users for pick-ups & \multicolumn{1}{c}{$\slash$} \\
   $N_d$ & Number of observed arrival vehicles for drop-offs & \multicolumn{1}{c}{$\slash$} \\
   $N_g$ & Number of GVST observations & \multicolumn{1}{c}{$\slash$} \\
   \hline
   \end{tabular}
   \label{tab:notations}
\end{table}

\subsection{Modeling the GVST}
\label{sec: GVST modeling}
 
In this section, we explain how the distribution of the random variable GVST within the proposed FQM is modeled.
Based on the generative process of GVST, we prove that deriving the distribution of GVST within the FQM is equal to modeling the sojourn time in the M/M/1/K queue with the first come, first served (FCFS) rule, as presented in Proposition~\ref{prop:gvst}. 

\begin{proposition}[Distribution of GVST]
\label{prop:gvst}
The distribution of GVST is identical to the distribution of sojourn time (also referred to as total customer waiting time) in the M/M/1/K queueing system with customer arrival rate $\lambda$, service rate $\mu$ and system capacity $K$, and follows the first come, first served (FCFS) rule. This equivalence holds when applying the FQM, which is characterized by a vehicle arrival rate $\lambda$, user arrival rate $\mu$ and unit capacity $K$.
\end{proposition}

\proof{
Recall that the vehicle arrival process in the FQM is modeled as the customer arrival process in the M/M/1/K queueing system, where vehicles arrive and wait for being picked up.
Correspondingly, the arrival users are modeled as the service process in the M/M/1/K queue, and the inter-arrival time of users corresponds to the service time of each customer.
According to Definition \ref{def:gvst}, the GVST can be interpreted as the waiting time of a vehicle from the moment it is dropped off until it is picked up under the assumption that the arrived user always picks up the vehicle arrived earliest.
This assumption exactly corresponds to the first come, first served (FCFS) rule in the M/M/1/K queue.
However, the GVST in the FQM does not directly correspond to the waiting time experienced by a customer in the queue of the M/M/1/K with FCFS rule. 
Specifically, the moment when a vehicle is picked up by a user in the FQM is analogous to a customer completing service and leaving the system in the M/M/1/K model. 
Similarly, the instant when a vehicle becomes the first vehicle in the queue corresponds to the moment a customer begins receiving service in the M/M/1/K queue. Therefore, the GVST can be decomposed into two components: (1) the time a vehicle spends waiting to become the first in the queue (equivalent to the customer waiting time in queue), and (2) the inter-arrival time before the subsequent user arrives to pick up the vehicle (corresponding to the service time in the M/M/1/K queue).
Consequently, the GVST in the FQM is equivalent to the sojourn time \citep{adan2002queueing} of a customer in the M/M/1/K queue, which encompasses both the waiting time in the queue and the service time. 
Thus, deriving the distribution of GVST in the FQM reduces to obtaining the distribution of the sojourn time in the M/M/1/K queue with FCFS rule.
\qed
}

\begin{figure}[h]
    \centering    \includegraphics[width=0.9\linewidth]{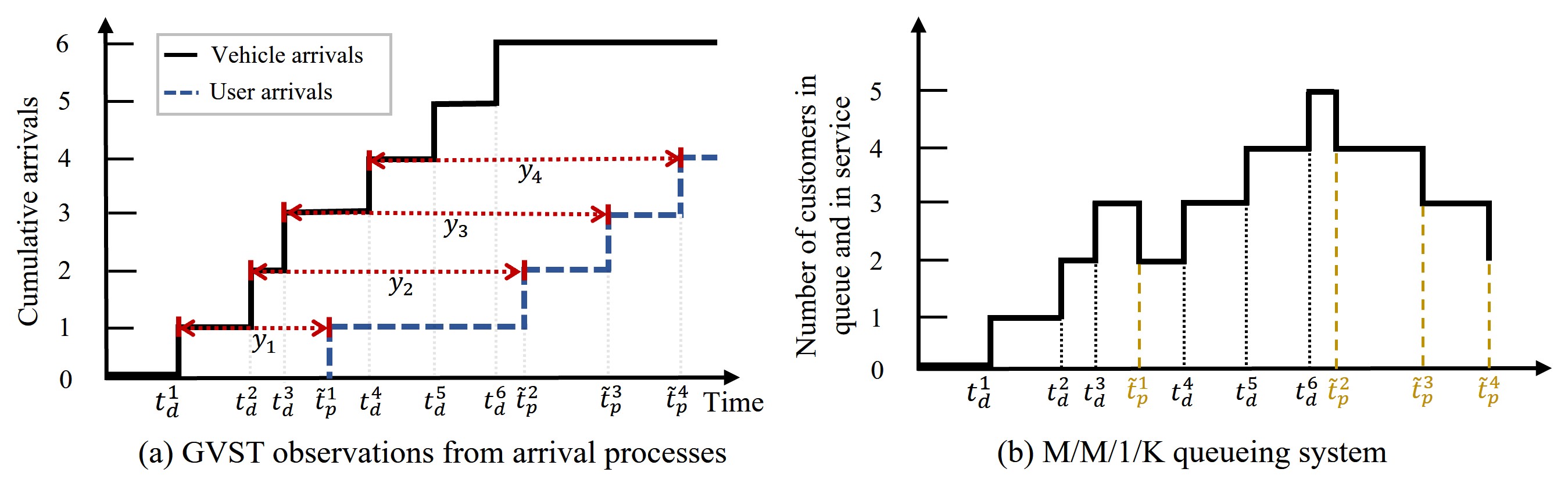}
    \caption{Example of observed arrival times.}
    \label{fig:BST}
\end{figure}

For a better understanding of the proposition, we analyze the example of arrival processes shown in Figure \ref{fig:BST}.
From the perspective of user and vehicle arrivals, the collection of GVST observations is illustrated in Figure \ref{fig:BST}(a). 
In this example, four GVST observations are obtained by computing $y_i=\tilde{t}^i_p-t^i_d$. 
From the queueing process perspective, the queue length process based on these arrivals is shown in Figure \ref{fig:BST}(b).
By examining the observations in Figure \ref{fig:BST}(b), it can be noted that the GVST in FQM corresponds to the time spent in the queue plus the time spent in service, which is referred to as sojourn times in the M/M/1/K queue. 
Specifically, each GVST observation can be decomposed into two parts, which correspond to the waiting time until it becomes the first vehicle in the queue and the inter-arrival time until the next user arrives to pick it up respectively. 
Taking $y_4$ in Figure \ref{fig:BST}(a) as an example, it can be divided into the first part $\tilde{t}^3_p - t^4_d$ and the second part $\tilde{t}^4_p - \tilde{t}^3_p$. 
The first part corresponds to the waiting time of customers in the queue in Figure \ref{fig:BST}(b), namely the waiting time the fourth arrived customer spends before the service begins.
The second part stands for the service time of the fourth customer in the M/M/1/K queue. 
If no other vehicles are available in the unit upon the arrival of a vehicle (e.g., in the case of $y_1$), the first component of its GVST is zero, as it is first in the queue at the time of arrival. 
This situation corresponds to the scenario in the M/M/1/K queue where a customer is served immediately upon arrival.

Based on Proposition~\ref{prop:gvst}, the probability density function (PDF) of the GVST can be derived by analyzing the corresponding sojourn time in a M/M/1/K queue with FCFS rule.
For the sake of mathematical tractability, we study the distribution when the queueing system is in a steady state.
Hence, all random variables discussed herein are considered under the assumption that the system has reached a steady-state condition.

The stationary M/M/1/K queue with FCFS rule is a classical model in queueing theory, with both its steady-state and transient probabilities well-studied \citep{alouf2001inferring, shortle2018fundamentals}. 
We briefly introduce some key formulations of M/M/1/K queue before deriving the closed-form expressions for the PDF of GVST. 
Let $P_x=Pr\{X=x\}$ denote the probability that the system is in state $x$, which indicates that the inventory is exactly $x$. 
In the context of the M/M/1/K queue, this denotes the probability of having exactly $x$ customers in the system. 
The birth-death process provides a straightforward method for determining the steady-state probabilities $P_x$ of the M/M/1/K queue. 
By modeling the system as a birth-death process, the steady-state probabilities can be derived using the balance equations and the normalization condition.
Details about the exact derivation of $P_x$ in various queueing settings can be found in the work of \citet{shortle2018fundamentals}. 
Here we directly present the steady-state probability of the inventory equal to $x$ can be computed in terms of the traffic intensity, denoted by $\rho$, as follows:
\begin{equation}
    P_x = \frac{(1-\rho)\rho^x}{1-\rho^{K+1}}, \; \text{where} \; \rho = \frac{\lambda}{\mu}.
    \label{eq:pn_rho}
\end{equation}

This study primarily focuses on scenarios in which the arrival rates are not equal, i.e., $\rho \neq 1$, as the case $\rho = 1$ is highly unlikely to occur in real-world settings and is discussed separately in \ref{App: A}.

Let $Y$ denote the r.v. GVST and $F_Y(y_i)$ represent the cumulative distribution function (CDF) of $i^{\mathrm{th}}$ GVST.
The elaborate derivation about the distribution of sojourn time in the M/M/1/K queue with FCFS rule can be found in \citet{sztrik2016basic} and \citet{shortle2018fundamentals}. 
Herein, we directly present the result of the distribution. 
Based on the law of total probability, the CDF of GVST can be derived as:

\begin{equation}
\begin{aligned} 
\label{eq:cdf}
F_Y(y_i) = \Pr\{Y \leq y_i\}
& = \sum_{x=0}^{K-1}\left(1-\sum_{z=0}^x\frac{(\mu y_i)^z}{z!}e^{-\mu y_i}\right)\frac{P_x}{1-P_K} \\
&=1-\sum_{x=0}^{K-1}\left(\sum_{z=0}^x\frac{(\mu y_i)^z}{z!}e^{-\mu y_i}\right)\frac{P_x}{1-P_K}.
\end{aligned} 
\end{equation}

Subsequently, the probability density function (PDF) of $i^{\mathrm{th}}$ GVST denoted by $f_Y(y_i)$ can be obtained by differentiating the CDF $F_Y(y_i)$ with respect to $y_i$:

\begin{equation}
\begin{aligned} 
\label{eq:pdf}
f_Y(y_i) = \frac{d}{dy} F_Y(y_i)
& = -\sum_{x=0}^{K-1} \frac{d}{dy_i} \left( \sum_{z=0}^x \frac{(\mu y_i)^z}{z!} e^{-\mu y_i} \right) \frac{P_x}{1-P_K} \\
&=\sum_{x=0}^{K-1} \mu \frac{(\mu y_i)^x}{x!} e^{-\mu y_i} \frac{P_x}{1-P_K} \\
&= \sum_{x=0}^{K-1}\frac{(\lambda y_i)^x}{x!}e^{-\mu y_i}\frac{(\mu - \lambda)\mu^K}{\mu^K - \lambda^K}.
\end{aligned} 
\end{equation}

\subsection{Real demand estimation}
\label{sec: MLE}

Once the distribution of GVST is specified with parameters, the best goodness of fit of the observed GVST can be achieved by searching for parameters of the FQM. 
In this manner, the exogenous process of the FQM, i.e. , user arrivals, can be estimated based on the observed endogenous process, namely the GVST observations.
To address this inverse queueing problem, the maximum likelihood estimation (MLE) method is utilized and estimate the user arrival rate $\mu$ characterizing the PDF of GVST. 
The parameter $K$ as the capacity of a unit is observable from the real-world data.
Based on the GVST observations $\boldsymbol{y}=\{y_i\}^{N_g}_{i=1}$, the likelihood function associated is conceptualized as the joint probability density function of $N_g$ observations. 
These observations are independent and identically distributed (i.i.d). 
The log-likelihood function is calculated through the product of the individual probability densities of each observation within the dataset.
Then, we define the natural logarithm of the likelihood function, $\mathcal{L}$ and obtain the following formula by substituting the PDF of GVST in Equation \ref{eq:pdf}:

\begin{equation}
\label{eq:likelihood}
\begin{aligned} 
\mathcal{L}(\lambda,\mu \mid \boldsymbol{y}, K) &= \log f_Y(y_1, \ldots, y_{N_g} \mid \lambda,\mu,K) \\
&=\sum_{i=1}^{N_g} \log f_Y(y_i \mid \lambda,\mu,K) \\
&= - \mu\sum_{i=1}^{N_g} y_i + N_g\log{\frac{\mu^{K+1}-\lambda \mu^K}{\mu^K - \lambda^K}}
+ \sum_{i=1}^{N_g}\log{\sum_{k=0}^{K-1}\frac{(\lambda y_i)^k}{k!}}.
\end{aligned} 
\end{equation}

The objective is to determine the optimal value of $\mu$ that maximizes the likelihood function, reflecting the probability of observed data.
Specifically, two distinct methods are proposed to estimate the exogenous process, which is the user arrival process. 
The first method, a two-sided estimation, simultaneously estimates both the supply side and the demand side.
The second method, a one-sided estimation, approximates the user arrival rate while treating the vehicle arrival process as observable.

\subsubsection{Two-sided estimation}

In this section, we introduce a two-sided method to estimate the supply side and the demand side of the FQM simultaneously, which correspond to the user arrival rate $\mu$ and vehicle arrival rate $\lambda$ respectively. 
The reason why we treat $\lambda$ as an unknown variable is that the censoring of demand for both pick-ups and drop-offs is found to exist when $\mu$ and $\lambda$ are close \citep{negahban2019simulation}. 
In other words, only when $\mu$ is considerably larger than $\lambda$, $\lambda$ can be viewed as a known variable. 
In other cases, treating the observed vehicle arrival rate as the real $\lambda$ will lead to bias in the model performance.
By applying the two-sided method, the MLE estimators of $\mu$ and $\lambda$ can become unbiased estimators when $N_g$ is sufficiently large.
To maximize the log-likelihood function $\mathcal{L}(\lambda, \mu | \boldsymbol{y}, K)$, we first derive the first-order partial derivative of $\mathcal{L}(\lambda, \mu | \boldsymbol{y}, K)$ concerning $\lambda$ and $\mu$ as

\begin{equation}
\frac{\partial \mathcal{L}}{\partial \lambda}= 
- \frac{N_g}{\mu-\lambda} + \frac{N_gK\lambda^{K-1}}{\mu^K-\lambda^K} 
+ \sum_{i=1}^{N_g}\frac{\sum_{k=0}^{K-1}\frac{y_i^k\lambda^{k-1}k}{k!}}{\sum_{k=0}^{K-1}\frac{(\lambda y_i)^k}{k!}},
\label{eq:partial lambda}
\end{equation}

\begin{equation}
\frac{\partial \mathcal{L}}{\partial \mu}= - \sum_{i=1}^{N_g} y_i + \frac{N_g}{\mu-\lambda} - \frac{N_gK\mu^{K-1}}{\mu^K-\lambda^K} + \frac{N_gK}{\mu}.
\label{eq:partial mu}
\end{equation}

Then we can acquire the Hessian matrix denoted by $H_t$ by computing the second derivatives of the log-likelihood function $\mathcal{L}(\lambda, \mu | \boldsymbol{y}, K)$.
\begin{align}
H_t &= \begin{bmatrix}
    \frac{\partial^2 \mathcal{L}}{\partial \mu^2} & \frac{\partial^2 \mathcal{L}}{\partial \mu \partial \lambda} \\
    \frac{\partial^2 \mathcal{L}}{\partial \lambda \partial \mu} & \frac{\partial^2 \mathcal{L}}{\partial \lambda^2} \\
\end{bmatrix} \nonumber \\
 & = \begin{bmatrix}
    - \frac{N_g}{(\mu - \lambda)^2} 
    - \frac{N_g K \lambda^K (\lambda^K - (1+K)\mu^K)}{\mu^2 (\mu^K-\lambda^K)^2}
    & \frac{N_g}{(\mu - \lambda)^2} - \frac{N_g K^2\mu^{K-1}\lambda^{K-1}}{(\mu^K-\lambda^K)^2} \\
    \frac{N_g}{(\mu - \lambda)^2} - \frac{N_g K^2\mu^{K-1}\lambda^{K-1}}{(\mu^K-\lambda^K)^2} 
    &  \text{\ding{72}}
\end{bmatrix},
\end{align}
where \ding{72} represents $\frac{-N_g}{(\mu-\lambda)^2} + \frac{N_g K\lambda^{K-2}(K\mu^K-\mu^K+\lambda^K)}{(\mu^K-\lambda^K)^2} + \sum_{i=1}^{N_g} \left(\frac{(-1 + K)^2 \Gamma(-1 + K, \lambda y_i)^2 y_i^2}{\Gamma(K, \lambda y_i)^2} + \frac{\Gamma(K) \Gamma(-2 + K, \lambda y_i) y_i^2}{\Gamma(-2 + K) \Gamma(K, \lambda y_i)}\right)$ and $\Gamma(\cdot)$ is the Gamma function.

Now we are ready to propose Proposition \ref{prop:two} indicating the two-sided estimation may have more than one solution.

\begin{proposition}
[Multiple solutions]
\label{prop:two}
The two-sided estimation problem can be formulated as:
\begin{equation}
    \hat{\lambda}, \hat{\mu} = \arg\max_{\lambda > 0, \mu > 0} l\left( \lambda, \mu | K,\boldsymbol{y} \right),
\end{equation}
where $\hat{\lambda}$ and $\hat{\mu}$ denote the estimators of $\lambda$ and $\mu$ respectively. 

Suppose that $\boldsymbol{y}$ and $K$ are observable, the estimators $\hat{\mu}$ and $\hat{\lambda}$ may have multiple solutions, implying the presence of multiple possible estimations that maximize the log-likelihood function.
\end{proposition}

\proof{See \ref{App: multi-solutions}}
\endproof

According to Proposition \ref{prop:two}, more than one local maxima exist for the maximization of likelihood $l(\lambda, \mu | K,\boldsymbol{y})$. 
Hopefully, we can use domain knowledge to refine the solution space and locate the solution we demand as presented in Proposition \ref{prop:two unique}.

\begin{proposition}
[Constrained uniqueness]
\label{prop:two unique}
We consider the partial derivatives of $l(\lambda, \mu | K,\boldsymbol{y})$ with respect to $\lambda$ (Equation \ref{eq:partial lambda}) and $\mu$ (Equation \ref{eq:partial mu}).
Setting these partial derivatives to zero and the system of equations can be obtained (See \ref{App:system of equation}).
The unique estimation can be derived by solving the system of equations, subject to the following constraints:

\begin{equation}
\begin{aligned}
    N_d \leq \hat{\lambda} \leq \lambda_{max} \\
    N_p \leq \hat{\mu} \leq \mu_{max},
\end{aligned}
\end{equation}
where $N_d$ denotes the observed demand for vehicle drop-offs computed by Equation \ref{eq:observed drop-offs}; $N_p$ denotes the observed demand for vehicle pick-ups computed by Equation \ref{eq:observed pick-ups}; $\lambda_{max}$ denotes the upper bound of $\lambda$ and $\mu_{max}$ represents the upper bound of $\mu$.

\end{proposition}

\proof{The system of equations can be solved using the least squares method \citep{hayashi2011econometrics}. 
The lower bounds of $\hat{\mu}$ and $\hat{\lambda}$ are established as the observed demand $N_p$ and $N_d$ making sure the estimations are efficient.
This is based on the premise that the real demand is definitely larger or equal to the observed demand.
The upper bounds $\mu_{max}$ and $\lambda_{max}$ are set as large quantities, which are dependent on the unit size and will be clarified in subsequent case studies.
The initial estimates for $\hat{\lambda}$ and $\hat{\mu}$ are set to $N_d$ and $N_p$. 
Incorporating these bounds reduces the set of potential solutions to a single unique estimation.
The resulting estimators correspond to the critical point closest to the observed demand, representing the efficient estimations of interest. \qed}

\subsubsection{One-sided estimation}

In this section, we introduce the one-sided method, which provides a closed-form estimation of the user arrival rate $\mu$. 
Compared to the two-sided method, the one-sided method requires one additional assumption that when estimating $\mu$, the vehicle arrival process is fully observed from the data, meaning that the value of $\lambda$ can be approximated by $N_d$ when the user arrival rate is larger than vehicle arrival rate, namely $\lambda \approx N_d,$ if $\mu > \lambda$. 
This assumption is reasonable because the demand for pick-up and demand for drop-off often follow opposite patterns \citep{negahban2019simulation}. 
As the ratio of $\mu$ to $\lambda$ increases, the observed vehicle arrival process increasingly approximates the real arrival process.
Then, the log-likelihood can be reformulated as
\begin{align*}
\mathcal{L}(\mu \mid \boldsymbol{y}, K, N_d) 
&= \sum_{i=1}^{N_g} \log f_Y(y_i \mid \mu, K, N_d) \\
&= - \mu \sum_{i=1}^{N_g} y_i 
   + N_g \log \left( \frac{\mu^{K+1} - N_d \mu^K}{\mu^K - {N_d}^K} \right)
   + \sum_{i=1}^{N_g} \log \left( \sum_{k=0}^{K-1} \frac{(N_d y_i)^k}{k!} \right).
\end{align*}

We aim to determine the optimal value of $\mu$ that maximizes the likelihood function, reflecting the probability of observed data. 
To this end, the first derivative of the log-likelihood $\mathcal{L}(\mu \mid \boldsymbol{y}, K, N_d)$ with respective to $\mu$ is calculated as

\begin{equation}
\frac{\partial \mathcal{L}(\mu \mid \boldsymbol{y}, K, N_d)}{\partial \mu}= -\sum_{i=1}^{N_g} y_i+\frac{N_g}{\mu-N_d} - \frac{N_g K\mu^{K-1}}{\mu^K-{N_d}^K} + \frac{N_g K}{\mu}.
\label{partial mu}
\end{equation}

Based on the first derivative $\frac{\partial \mathcal{L}}{\partial \mu}$, we can compute the Hessian matrix ($H_o$) of the log-likelihood as:

\begin{equation}
H_o = \frac{\partial^2 \mathcal{L}}{\partial \mu^2}
= \frac{-N_g}{(\mu - {N_d})^2} 
+ \frac{N_g K{N_d}^{K}((1+K)\mu^K-{N_d}^K)}{\mu^2(\mu^K-{N_d}^K)^2}. 
\label{one hessian}
\end{equation}

Then, we demonstrate that when the GVST $\boldsymbol{y}$, the capacity of the unit $K$, and the vehicle arrival rate $\lambda$ are observable, it is theoretically sufficient to uniquely estimate $\mu$, as established in Proposition~\ref{prop:id one}. Such an identifiability condition ensures that, under the given assumptions, there exists a unique value of $\mu$ that maximizes the likelihood function, thus allowing for the precise estimation of $\mu$ from the GVST observations.

\begin{proposition}
[Identifiability]
\label{prop:id one}
The parameter $\mu$ can be estimated by finding the value $\hat{\mu}$ that maximizes $\mathcal{L}(\mu \mid \boldsymbol{y}, K, N_d)$, which is formulated as

\begin{equation}
    \hat{\mu} = \arg\max_{\mu > 0} \mathcal{L}(\mu \mid \boldsymbol{y}, K, N_d).
\end{equation}

The objective function $\mathcal{L}(\mu \mid \boldsymbol{y}, K, N_d)$ is concave when $\mu > N_d$, meaning that $\mu$ can be uniquely estimated from the observed $\boldsymbol{y}$, $K$, and $N_d$.
\end{proposition}

\proof{
Focusing on the overall sign of Equation \ref{one hessian}, the two fractions are combined, and the terms that are known to be positive are eliminated. Additionally, the parameter $\rho=\frac{{N_d}}{\mu}$ is introduced to substitute for ${N_d}$ and $\mu$. 
Consequently, the expression can be reformulated as $f = K^2(\rho-1)^2\rho^K-K(\rho-1)^2\rho^K(\rho^K-1)-(\rho^K-1)^2$. 
Considering that $\rho<1$ and that $K$ is a relatively large integer, both the first and second terms of $f$ approach 0 as $\rho$ decreases from 1 to 0 due to the rapid decay of $\rho^K$. Therefore, it follows that $f \approx -1<0$. As a result, the Hessian matrix is negative semidefinite, and the function $l(\mu | \lambda, K,\boldsymbol{y})$ is concave when $\rho<1$. This implies that $\mu$ can be uniquely estimated, thereby validating the proposition. \qed}

As is presented in Proposition \ref{prop:id one}, the estimator $\hat{\mu}$ that maximizes $\mathcal{L}(\mu \mid \boldsymbol{y}, K, N_d)$ can be computed by setting Equation \ref{partial mu} to zero.
Then we have the following equation:

\begin{equation}
-\sum_{i=1}^{N_g}y_i +\frac{N_g}{\mu-{N_d}} - \frac{N_g K\mu^{K-1}}{\mu^K-{N_d}^K} + \frac{N_g K}{\mu} = 0.
\label{eq:0}
\end{equation}

Now we present Proposition~\ref{prop:close form}, in which the real demand can be analytically derived in closed form when the demand for pick-ups exceeds the demand for drop-offs.

\begin{proposition}
[Closed-form solution]
\label{prop:close form}
The arrival rate of users, $\mu$ can be approximated by the closed-form expression when $\mu > N_d$ as:
\begin{equation}
\tilde{\mu} = N_d + \frac{N_g}{\sum_{i=1}^{N_g}y_i},
\label{eq:closed form}
\end{equation}
where $\tilde{\mu}$ is the approximated estimator of $\mu$.
\end{proposition}

\proof{
Since obtaining an exact solution for $\mu$ from Equation \ref{eq:0} is analytically intractable, an approximate solution can be determined by introducing the traffic intensity $\rho=\frac{{N_d}}{\mu}$.
Under this substitution, the third term of Equation \ref{eq:0} is transformed into $\frac{N_g K\mu^{-1}}{1-\rho^K}$.
When $\rho < 1$, it is reasonable to approximate this term by $N_g K\mu^{-1}$. 
This approximation is valid because $\rho^K$ becomes very small and negligible as $K$ increases, noting that $K$ typically represents a unit capacity and is usually a large integer.
This yields the following expression: $\sum_{i=1}^{N_g}y_i -\frac{N_g}{\mu-{N_d}} + N_g K\mu^{-1} - \frac{N_g K}{\mu} \approx 0$.
By consolidating the terms that involve $\mu$, we obtain the approximated estimator $\tilde{\mu} = {N_d} + N_g/ \sum_{i=1}^{N_g}y_i$.
This completes the derivation. \qed}

Deriving the closed-form solution for $\mu$ holds significant value, as it simplifies an otherwise complex relationship, thus enabling more tractable analysis and interpretation. 
This closed-form expression provides a streamlined approach for practical applications in real demand estimation. 
The importance of this simplification lies in its ability to transform what would be a computationally demanding task into one that can be readily applied to real-world scenarios.
It is worth noting that the approximated solution is derived under the condition $\mu>\lambda$, which corresponds to situations where the demand for pick-ups exceeds that of drop-offs.
This condition reflects most cases in realistic scenarios, where the real demand is censored due to the higher demand for picking up vehicles. 
Therefore, by utilizing the closed-form solution, decision-makers can obtain a simple yet effective rule of thumb for anticipating real demand estimation. 

While Proposition~\ref{prop:close form} provides a convenient closed-form approximation for the estimation of $\mu$, it is important to recognize that the estimator $\tilde{\mu}$ exhibits a systematic bias, tending to overestimate the user arrival rate $\mu$. This effect is formalized in the proposition \ref{prop:overestimation}.

\begin{proposition}
[Overestimation]
\label{prop:overestimation}
The closed-form estimator $\tilde{\mu}$ defined in Proposition~\ref{prop:close form} generally overestimates the true $\mu$, i.e., $\tilde{\mu} > \mu$ for the underlying system.

\end{proposition}
\proof{
The closed-form solution in Equation~\eqref{eq:closed form} is derived under the approximation that the term $\rho^K$ in the denominator of the third term of Equation~\eqref{eq:0} is negligible when $K$ is large, i.e., $1-\rho^K \approx 1$. This approximation systematically ignores the small but nonzero contribution of $\rho^K$, which otherwise slightly reduces the denominator, thus increasing the value of the term $\frac{N_g K\mu^{-1}}{1-\rho^K}$ compared to its approximation $N_g K\mu^{-1}$. As a result, the closed-form estimator $\tilde{\mu}$ must compensate by yielding a higher value than the true $\mu$ to satisfy the approximate balance equation. 
Formally, for any finite $K$ and $\rho<1$, we have $1-\rho^K < 1$, which implies $\frac{1}{1-\rho^K} > 1$. Therefore, the use of $N_g K\mu^{-1}$ as an approximation instead of $\frac{N_g K\mu^{-1}}{1-\rho^K}$ leads to a systematic underestimation of the overall value, which is reflected as $\sum_{i=1}^{N_g}y_i -\frac{N_g}{\mu-{N_d}} + N_g K\mu^{-1} - \frac{N_g K}{\mu} < 0$.
Therefore, the relationship between $\tilde{\mu}$ and $\mu$ is represented as $\mu < N_d + \frac{N_g}{\sum_{i=1}^{N_g}y_i} = \tilde{\mu}$, which means the closed-form solution $\tilde{\mu}$ tends to overestimate $\mu$.  
\qed}

\subsection{Solution algorithms}

We summarize the pseudocode of the two-sided and one-sided estimation process, which is presented in Algorithm \ref{alg:algorithm}. 
Given the observed data $\mathcal{D} = \{(I_n, t^*_n, k_n)\}_{n=1}^N$, the estimator $\hat{\mu}$ is obtained when applying the two-sided method while the estimator $\tilde{\mu}$ is derived using the one-sided method.

\begin{algorithm}[h]
\flushleft
\caption{: Estimation procedures.}
\label{alg:algorithm}
\begin{algorithmic}
\REQUIRE observed data $\mathcal{D} = \{(I_n, t^*_n, k_n)\}_{n=1}^N$ and vehicle capacity of the unit $K$.
\end{algorithmic}
\begin{algorithmic}[1]
\STATE Compute the sequence of observed drop-off times $\boldsymbol{t_d}$ and total drop-off number $N_d$ based on Equation \ref{eq:observed drop-offs}; 
Compute the sequence of observed pick-up times $\boldsymbol{\tilde{t}_p}$ and total drop-off number $N_p$ based on Equation \ref{eq:observed pick-ups}.
\STATE Compute the observations of GVST $\boldsymbol{y}=\{y_i\}^{N_g}_{i=1}$ by applying the swap scheme based on Equation \ref{eq:gvst}.
\STATE Select the estimation method: \textbf{two-sided} or \textbf{one-sided};
\IF{method == two-sided}
    \STATE Initialize the parameters $\lambda$ and $\mu$ as $N_d$ and $N_p$.
    \WHILE{changes of $\hat{\lambda}$ and $\hat{\mu}$ are greater than tolerances}
    \STATE Construct the system of equations using Equation \ref{eq:partial lambda} and \ref{eq:partial mu}.
    \STATE Solve for $\Delta \theta = \begin{pmatrix}\Delta\hat{\lambda} \\ \Delta\hat{\mu}\end{pmatrix}$ using the Least Squares method based on observations of GVST $\boldsymbol{y}$ and parameter $K$.
    \STATE Update parameters by $\hat{\lambda} + \Delta\hat{\lambda}$ and $\hat{\mu} + \Delta\hat{\mu}$.
    \ENDWHILE
    \STATE \textbf{return} the estimator $\hat{\mu}$ and $\hat{\lambda}$.
\ELSE
    \STATE Initialize the parameter $\mu$ as $N_p$.
    \STATE The closed-form estimator $\tilde{\mu}$ can be derived based on Equation \ref{eq:closed form}.
    \STATE \textbf{return} the estimator $\tilde{\mu}$.
\ENDIF

\end{algorithmic}
\end{algorithm}

\section{Numerical Experiments}
\label{sec:Experiments}

Following the previous research, we validate the proposed method by executing experiments on synthetic datasets \citep{gammelli2020estimating, huttel2022modeling, wang2023recovering, wang2023demand} and analyze the estimated results on the real-world datasets from bike-sharing and shared e-scooter systems. 
The reason is that the real demand for micro-mobility services is intractable. 
By applying simulation experiments to generate the synthetic datasets, the real demand can be obtained for validation and comparative analysis.

\subsection{Synthetic dataset}

In this section, we conduct simulation experiments to generate a synthetic dataset that is used for validating the performance of the proposed estimation methods.

\subsubsection{Simulation experiments for validation}

First, we design simulation experiments to generate synthetic datasets to validate our proposed method. 
The same discrete event simulation model as in \citet{negahban2019simulation} is applied to simulate the operation of a unit of shared micro-mobility services.
Aligning with previous studies \citep{o2015data, wang2023demand}, we consider Poisson user and vehicle arrivals.
The time period $T$ for the simulation is set as one hour \citep{negahban2019simulation, wang2023recovering}.
A user will abort if there is no vehicle available. Similarly, vehicles leave the unit if the capacity is full. 
The vehicle pick-up and drop-off time is negligible, i.e., cost zero simulation time.
Table \ref{tab:simulation} shows the details of the parameters used in the simulation experiments. 
By changing the configurations of user arrival rates, various scenarios are simulated to cover the possible conditions in the real world. 
As the user arrival rate $\mu$ increases, the supply of vehicles will gradually become unable to meet the growing demand for pick-ups, resulting in a mismatch between the observed demand and the real demand. 
In this manner, the observed data recording the time epochs of successful pick-up and drop-off events by users and vehicles are generated. 
For each preset $\mu$, observed dataset $\mathcal{D} = \{(I_n, t^*_n, k_n)\}_{n=1}^N$ is collected, and the GVST observations can be computed based on Equation \ref{eq:gvst}. 
The simulation process is conducted 200 times to generate independent samples of the dataset $\mathcal{D}$.

\begin{table}[ht]
    \centering
    \footnotesize
    \caption{Parameter choices for the simulation experiments}
    \begin{tabular}{lc}
     \toprule
     Parameters & Value or range of parameters \\
     \hline
     Vehicle arrival rate $\lambda$ & 100 people/hour\\
     User arrival rate $\mu$ & 105 to 200 in 10 increments (people/hour) \\
     Dock capacity & 20 vehicles \\
     Number of GVST observations & 5000 \\
     Number of replications & 200 \\
     \bottomrule
    \end{tabular} 
    \label{tab:simulation}
\end{table}

Then, the proposed methods are applied to estimate the arrival rate of users arrival $\mu$. 
To avoid achieving unrealistic results (extremely large or negative values), we constrain the parameter space to make sure the estimations conform to real-world expectations. 
Based on the observed data $\mathcal{D}$, we compute the observed number of pick-ups $N_p$ and drop-offs $N_d$ based on Equation \ref{eq:observed drop-offs} and Equation \ref{eq:observed pick-ups}. The lower bounds of $\hat{\lambda}$ and $\hat{\mu}$ are set as $N_d$ and $N_p$ respectively. The upper bounds of $\lambda_{max}$ and $\mu_{max}$ are selected as ten times the highest $N_d$ and $N_p$ observed historically.

\subsubsection{Model performance}

We implement the proposed method on the synthetic dataset to estimate the real demand. 
Three primary performance metrics, Root Mean Square Error (RMSE), Mean Absolute Percentage Error (MAPE), and Mean Absolute Error (MAE) are employed to evaluate the effectiveness of our model. 
These evaluation metrics provide a comprehensive assessment of the estimated accuracy, each highlighting different aspects of errors.
To better evaluate the performance of the two-sided and one-sided methods, we introduce a baseline method (BL) as an extension of the one-sided estimation. 
As discussed in Proposition~\ref{prop:close form}, the one-sided estimation is a closed-form approximation to the solution of Equation \ref{eq:0}. 
While the estimator in the BL method, denoted as $\hat{\mu}_{BL}$, is obtained by solving Equation \ref{eq:0} using a numerical algorithm, the Newton-Raphson method. 
To put it simply, the one-sided estimator $\tilde{\mu}$ serves as a computationally efficient and analytically tractable approximation of the BL estimator $\hat{\mu}_{BL}$.
Furthermore, the $\hat{\mu}_{BL}$ of the BL method can be regarded as an approximation to $\hat{\mu}$ by assuming that the $\lambda$ is fully observable from data.

\begin{table}[ht]
    \centering
    \footnotesize
    \caption{Performance of the proposed estimation methods on synthetic data}
    \begin{tabular}{llllllllllllllllll}
     \toprule
     $\lambda$ & $\mu$ & $K$ & $\hat{\mu}$ & \multicolumn{3}{c}{Two-sided} & \multicolumn{1}{c}{$\tilde{\mu}$} & \multicolumn{3}{c}{One-sided} & \multicolumn{1}{c}{$\hat{\mu}_{BL}$} & \multicolumn{3}{c}{BL}\\
     \cline{5-7} \cline{9-11} \cline{13-15}  & & & & RMSE & MAPE & MAE & & RMSE & MAPE & MAE & & RMSE & MAPE & MAE\\
    \hline
100 & 105 & 20 & 106.33 & 1.97 & 2.86\% & 3.00 & 109.22 & 2.12 & 4.02\% & 4.22 & 102.91 & 2.03 & 3.34\% & 2.45\\
100 & 115 & 20 & 116.31 & 1.88 & 1.96\% & 2.26 & 117.18 & 1.97 & 2.31\% & 2.65 & 114.11 & 1.93 & 2.15\% & 2.30\\
100 & 125 & 20 & 125.88 & 1.84 & 2.07\% & 2.59 & 126.28 & 1.99 & 2.18\% & 2.67 & 125.08 & 1.86 & 2.13\% & 2.61\\
100 & 135 & 20 & 135.58 & 1.88 & 2.04\% & 2.76 & 135.75 & 1.94 & 2.23\% & 2.70 & 135.30 & 1.92 & 2.14\% & 2.82\\
100 & 145 & 20 & 145.44 & 1.79 & 1.71\% & 2.48 & 145.50 & 1.70 & 1.69\% & 2.31 & 145.33 & 1.71 & 1.60\% & 2.31\\
100 & 155 & 20 & 155.26 & 1.91 & 1.84\% & 2.85 & 155.45 & 1.85 & 1.73\% & 2.69 & 155.39 & 1.85 & 1.74\% & 2.70\\
100 & 165 & 20 & 165.42 & 1.83 & 1.61\% & 2.66 & 165.72 & 1.78 & 1.53\% & 2.52 & 165.70 & 1.78 & 1.53\% & 2.52\\
100 & 175 & 20 & 175.19 & 1.87 & 1.62\% & 2.83 & 174.97 & 1.87 & 1.61\% & 2.82 & 174.57 & 2.52 & 1.82\% & 3.19\\
100 & 185 & 20 & 185.31 & 1.90 & 1.62\% & 3.00 & 185.39 & 1.89 & 1.61\% & 2.98 & 185.38 & 1.89 & 1.61\% & 2.98\\
100 & 195 & 20 & 194.99 & 1.93 & 1.57\% & 3.06 & 195.14 & 1.97 & 1.59\% & 3.08 & 195.12 & 1.95 & 1.62\% & 3.12\\
     \bottomrule
    \end{tabular} 
    \footnotesize
      \item \hspace*{35em} \footnotesize{\textit{Note:} RMSE and MAE are in (people / hour).}
    \label{tab: proposed performance}
\end{table}

The performances of the two-sided method, one-sided method, and BL method on the synthetic dataset are summarized in Table \ref{tab: proposed performance}. In this table, $\hat{\mu}$, $\tilde{\mu}$ and $\hat{\mu}_{BL}$ denote the expected values of the estimators using the respective methods over 200 independently generated data $D$ generated for each combination of $\mu$ and $\lambda$. 
By comparing the performances of these methods, we found that when the $\mu$ is significantly greater than $\lambda$ ($\mu \geq \lambda+15$), both two-sided and one-sided methods can achieve good results. 
However, when the $\mu$ and $\lambda$ are close to each other ($\mu \leq \lambda+15$), the estimations from the one-sided method and BL method deviate from the true value. 
On the contrary, the estimation $\hat{\mu}$ from the two-sided method is very close to the true value. 
This result indicates that, on average, the two-sided method is almost unbiased and demonstrates lower bias compared to both the one-sided method and the BL method.
The two-sided method is likely to give estimations that are consistently centered around the true value across the 200 trials, which suggests good accuracy of the estimations.
Nonetheless, the two-sided method does not significantly outperform the one-sided method and BL method across all three metrics. 
This implies that the two-sided method can obtain results with a lower bias, but could have higher variability in the estimates across different trials. 
This variability may result in marginally higher percentage error in individual estimates.
Nevertheless, the two-sided method is preferred due to the lower bias and similar performances on the metrics, especially the absolute scale of demand estimates is crucial. 
On the other hand, it can be seen that the performance of the closed-form solution improves as the user arrival rate $\mu$ increases. 
The reason is that the closed-form solution tends to converge to the true solution with $\rho=\lambda/\mu$ approaching zero. 
As expected, the performance of closed-form estimations closely aligns with the $\hat{\mu}_{BL}$ of the BL method. 
Although the performance is slightly inferior to that of the two-sided method, the one-sided method still achieves highly accurate estimation results especially when $\mu > \lambda+25$, which demonstrates its efficiency. 
Therefore, the closed-form solution can serve as an efficient approach for recovering real demand, particularly in situations where supply is severely insufficient to meet all demand.

\begin{figure}[h]
    \centering    \includegraphics[width=0.8\linewidth]{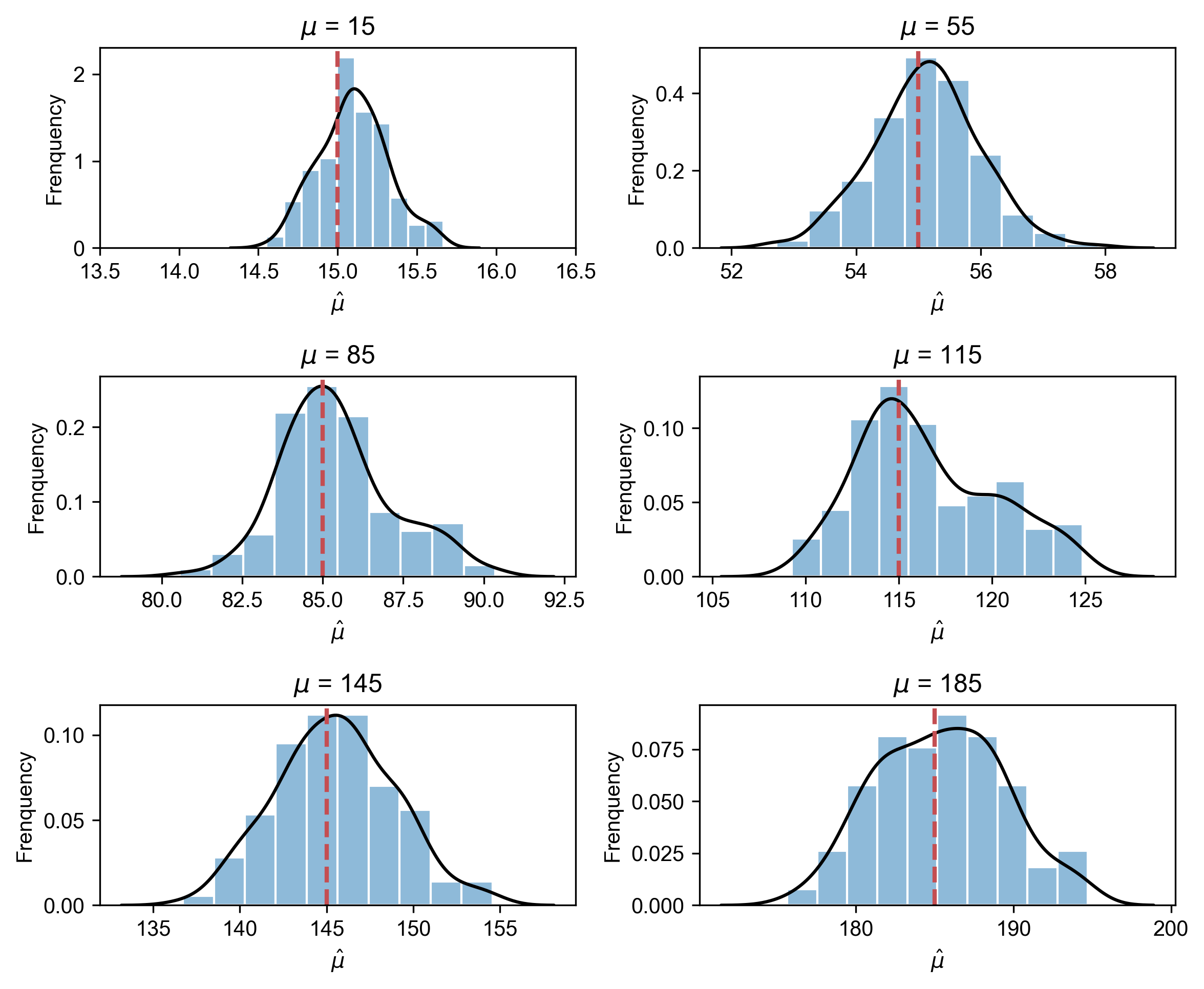}
    \caption{Histogram of $\hat{\mu}$ on synthetic dataset.}
    \label{fig:distribution}
\end{figure}

To evaluate the robustness of the proposed two-sided and one-sided methods, we conduct two sensitivity analyses that are designed to assess the impact of critical factors on the estimation results, ensuring the reliability and generalizability of our approaches.
The first sensitivity analysis examining the influence of different capacities of the unit on the outcomes is presented in \ref{App:SA_K}. 
The second analysis investigating the effects of varying GVST observation numbers on the estimation results is shown in \ref{App:SA_GVST}. 
As demonstrated in Proposition~\ref{prop:overestimation}, the one-sided estimation method tends to generate overestimated results. 
To further investigate this phenomenon, we analyze the distribution of the two-sided estimator $\hat{\mu}$ under various settings of $\mu$ in Figure \ref{fig:distribution}
It can be seen that the estimated $\hat{\mu}$ asymptotically follows a Gaussian distribution. 
This implies that the MLE achieves desirable consistency in the estimation procedure. If the sample size becomes larger and grows to infinity, the MLE estimator will converge in probability to the true parameter value. 
The mean of these parameter estimates is proximate to the true values, indicating that our estimator is unbiased on average. 
The desirable variance of the estimates also demonstrates the effectiveness of our estimation method.

To summarize, the closed-form one-sided estimation demonstrates comparable efficiency to the two-sided method and BL method, as evidenced by similar performance outcomes. 
Furthermore, the two-sided estimation exhibits greater reliability than the one-sided method and BL method, particularly when the values of $\mu$ and
$\lambda$ are proximate. 
Therefore, this study highlights the two-sided estimation and closed-form one-sided estimation to recover the real demand.

\subsubsection{Comparative analysis}

We further compare the proposed estimation methods with two state-of-the-art methods to demonstrate the efficiency. The first benchmark model is the Censored Gaussian Processes (CGP) proposed by \citet{gammelli2020estimating}.
The CGP assumes a Gaussian distribution for the real demand and considers the censoring mechanism by incorporating it into the likelihood function.
The second model models the true mobility demand via Multi-Output Censored Quantile Regression Neural Networks (Multi-CQRNN) \citep{huttel2022modeling}. 
Unlike traditional Tobit regression, which focuses on the mean of the dependent variable, Multi-CQNN aims to estimate conditional quantiles. This allows for a more detailed understanding of the distribution of the dependent variable.
The methods proposed in \citet{wang2023demand} and \citet{wang2023recovering} are not compared because external data are required to estimate the real demand. However, the influential factors \citep{wang2023demand} and migration behaviours \citep{wang2023recovering} are not needed to generate the real demand in our cases.

The input data sample of these baseline models is represented as $(x^*, l, y^*)$, where $x^*$ is the explanatory features, $l$ is the binary censorship label and $y^*$ is the censored observed demand of consecutive periods. To generate the above data structure, we consider the time period as 15 minutes \citep{gammelli2020estimating} and aggregate the observed demand in each period as $y^*$. 
If at any moment during this period there were no vehicles available at the unit, then $l$ equals 1; otherwise $l$ equals 0. 
The covariates $x^*$ for CGP are the temporal index, while the $x^*$ for CQRNN is set as the observed demand of the past eight periods.
In this manner, our synthetic dataset can be transformed into a dataset $\mathcal{D}=\{(x^*_i, l_i, y^*_i)\}_{i=1}^{N_t}$, where $N_t$ represents the number of time interval, and then used to validate the CGP and CQRNN.

\begin{table}[h]
    \centering
    \footnotesize
    \caption{Comparison of the proposed model and other state-of-the-art models.}
    \begin{tabular}{lccccccccccccccc}
    \toprule 
    $\mu$ & \multicolumn{3}{c}{One-sided} & & \multicolumn{3}{c}{Two-sided} & & \multicolumn{3}{c}{CGP} & & \multicolumn{3}{c}{CQRNN}\\
     \cline{2-4} \cline{6-8} \cline{10-12} \cline{14-16} &  RMSE & MAPE & MAE & & RMSE & MAPE & MAE & & RMSE & MAPE & MAE & & RMSE & MAPE & MAE \\
    \hline
    95 & 2.36 & 5.69\% & 5.40 & & \textbf{1.73} & \textbf{2.56}\% & 2.44 & & 
    3.45 & 3.27\% & 2.97 & & 3.10 & 2.72\% & \textbf{2.42} \\
    105 & 2.12 & 4.02\% & 4.22 & & \textbf{1.97} & \textbf{2.86}\% & 3.00 & & 
    3.59 & 4.43\% & 3.08 & & 3.27 & 3.85\% & \textbf{2.94} \\
    115 & 1.97 & 2.31\% & 2.65 & & \textbf{1.88} & \textbf{1.96}\% & \textbf{2.26} & & 
    5.28 & 5.02\% & 4.62 & & 3.67 & 4.68\% & 3.73 \\
    125 & 1.99 & 2.18\% & 2.67 & & \textbf{1.84} & \textbf{2.07}\% & \textbf{2.59} & & 
    5.91 & 5.25\% & 5.32 & & 4.39 & 4.97\% & 3.98 \\
    135 & 1.94 & 2.23\% & \textbf{2.70} & & \textbf{1.88} & \textbf{2.04}\% & 2.76 & & 
    6.22 & 5.28\% & 5.77 & & 5.76 & 5.17\% & 4.29 \\
    145 & \textbf{1.70} & \textbf{1.69}\% & \textbf{2.31} & & 1.79 & 1.71\% & 2.48 & & 
    7.27 & 5.89\% & 7.09 & & 6.79 & 5.41\% & 6.40 \\
    155 & \textbf{1.85} & \textbf{1.73}\% & \textbf{2.69} & & 1.91 & 1.84\% & 2.85 & & 
    9.58 & 6.50\% & 8.54 & & 8.64 & 5.56\% & 5.01 \\
    \bottomrule
    \end{tabular}    
    \footnotesize
    \item \hspace*{37em} \footnotesize{\textit{Note:} RMSE and MAE are in (people / hour).}
    \label{tab:baselines}
\end{table}

The estimation results of real demand using the three methods are presented in Table \ref{tab:baselines}. It can be noted that the proposed method is superior to the baseline models by achieving lower values of the three metrics. 
The baseline models are able to recover the real demand and achieve satisfactory results when the arrival rates of users and vehicles are close.
However, there will be a decline in the performance with the increase of $\mu$. 
This phenomenon may be attributed to the limitations of covariates, particularly the censoring present in historical observations.
Such a problem results in insufficient information about the real demand captured within the covariates, thereby diminishing the accuracy of demand estimation.
In our case, the censoring pattern is further complicated by bilateral arrival processes, rendering the Tobit-based likelihood approach intractable. 
Moreover, the assumption made by Tobit models that the censoring mechanism is independent of the latent variable (i.e., the real demand) may be violated, as the probability of censoring increases with higher levels of real demand.
For CQRNN, estimation becomes challenging and less precise when the censoring rate (i.e., the proportion of censored samples) approaches or exceeds the quantile of interest, namely the mean.
The results indicate that censored modeling approaches can only partially recover the real demand, rather than fully reveal its underlying distribution.
The reason is that purely relying on censored models fails to account for the generative process underlying the gap between observed and real demand, making accurate estimation of real demand intractable.
To summarize, the two proposed estimation methods are generally more reliable and accurate than other models in terms of overall performance.
In the following two sections, numerical experiments of applying the proposed two-sided estimation method on real-world datasets are conducted to demonstrate its effectiveness empirically.
Both dock-based bike-sharing and dockless e-scooter systems are included and analyzed. 

\begin{figure}[h]
    \centering    \includegraphics[width=0.9\linewidth]{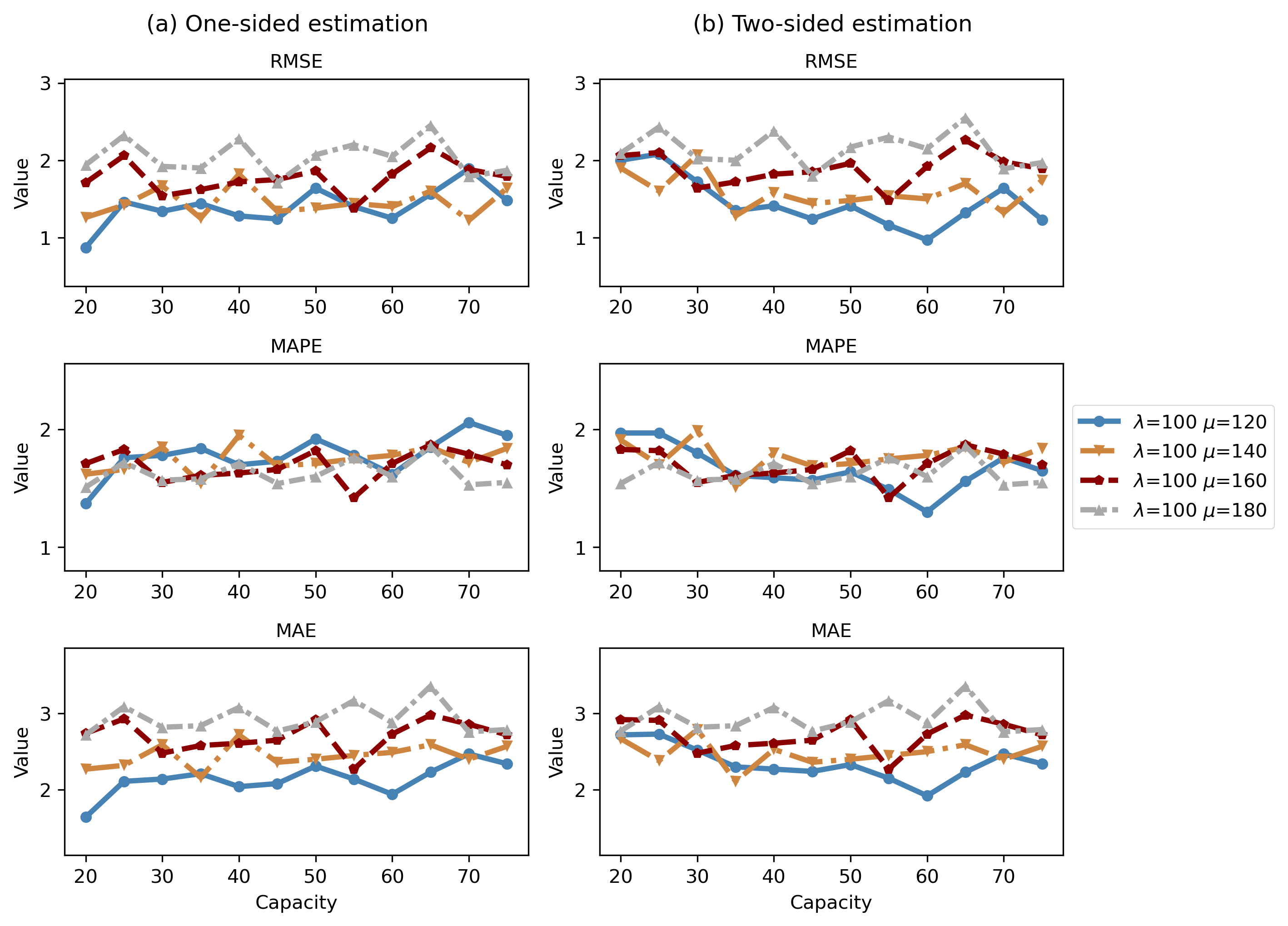}
    \caption{Sensitivity analysis regarding unit capacity.}
    \label{fig:SA K}
\end{figure}

\subsubsection{Sensitivity analysis regarding unit capacity}
\label{App:SA_K}

We conduct the sensitivity analysis to examine the effect of unit capacity on the estimation result. A larger unit capacity signifies a larger scale, enabling the simultaneous accommodation of more vehicles.
We vary the unit capacity (in vehicles) in increments of 5, ranging from 20 to 80 as a case study, to assess its impact on the estimation results.
The results of both two-sided and one-sided methods are shown in Figure \ref{fig:SA K}.
For each curve in the figure, the x-axis represents the unit capacity and the y-axis represents the performance of the estimation results on the three metrics. One can see that the performance of our methods tends to fluctuate randomly with respect to the changes in capacities.
The relatively stable fluctuations of results indicate that the proposed method is robust to changing unit capacity.

\begin{figure}[h]
    \centering    \includegraphics[width=0.9\linewidth]{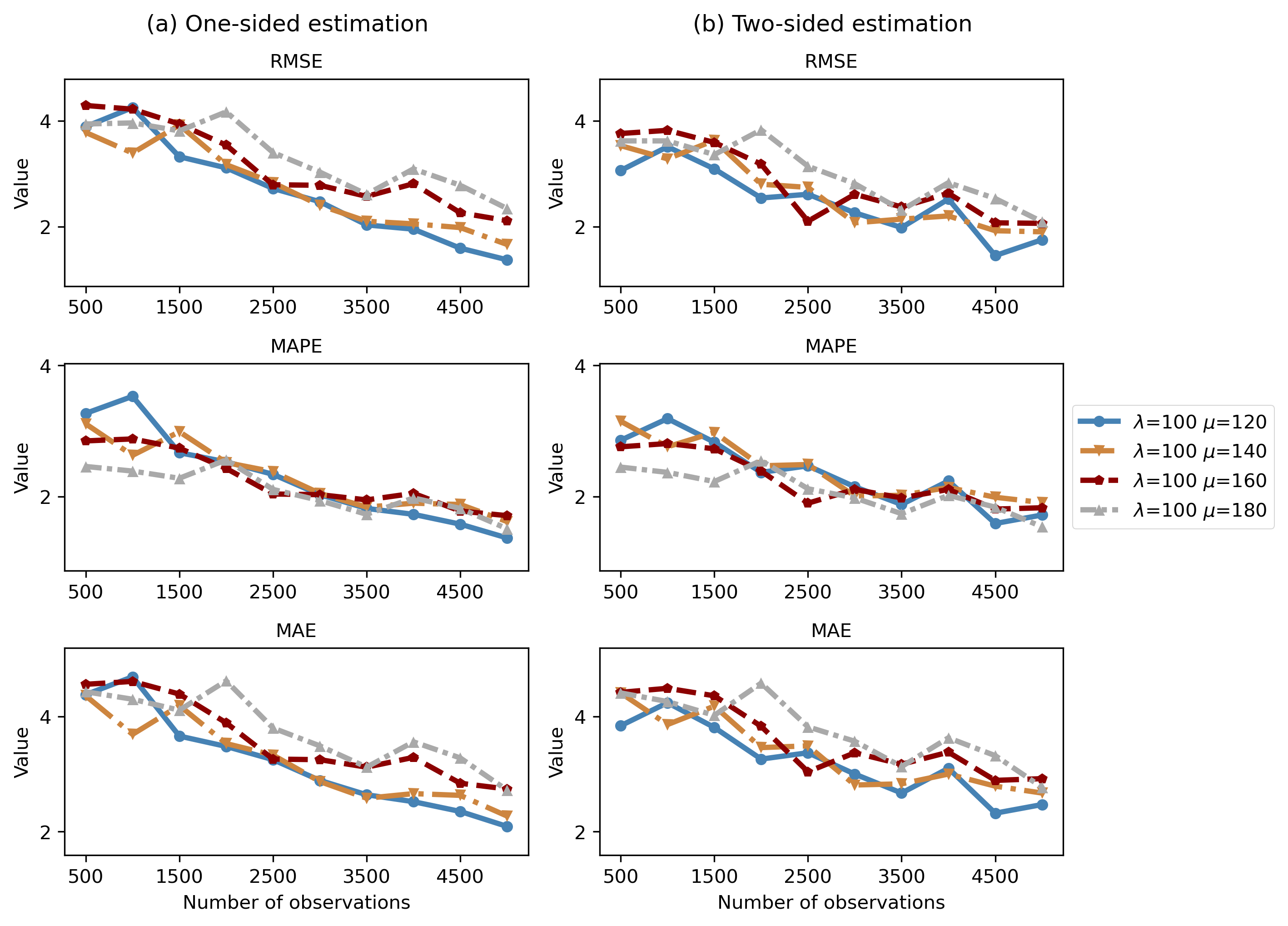}
    \caption{Sensitivity analysis regarding the number of GVST observations.}
    \label{fig:SA GVST}
\end{figure}

\subsubsection{Sensitivity analysis regarding the number of GVST observations}
\label{App:SA_GVST}

Subsequently, we perform a sensitivity analysis to determine how the number of GVST observations impacts the estimation results.
The number of GVST observations in a unit is affected by the user arrival rate $\mu$ and vehicle arrival rate $\lambda$ and the unit with more pick-ups and drop-offs will generate more observations of GVST.
The result for varying the number of GVST observations from 500 to 5000 in 500 increments for estimation is shown in Figure \ref{fig:SA GVST}.
It can be noted that the performance exhibits a significant enhancement as the number of GVST observations increases.
The reason is that MLE is a statistically consistent estimator, meaning that as the sample size grows, the estimates produced by the model converge to the true parameter values. Larger datasets provide more information, reducing the variance of the estimators and leading to more precise parameter estimates.
Note that the model performs acceptably even when the number of GVST observations is 500, with the MAPE remaining below $4\%$ and the MAE below $5$ people/hour.

\begin{figure}[h]
    \centering
    \includegraphics[width=0.8\linewidth]{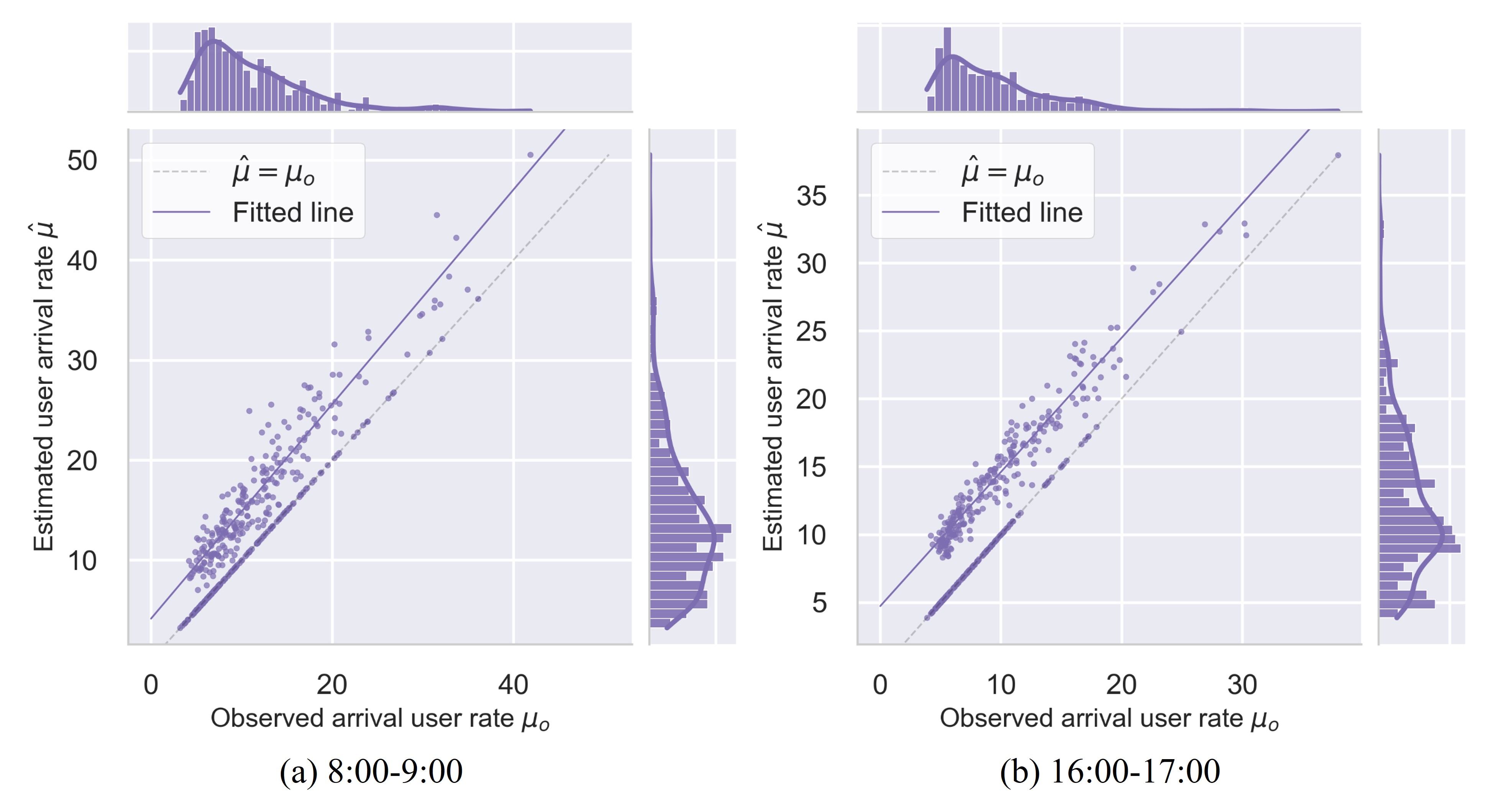}    \caption{Distribution of MLE estimators.}
    \label{fig:bike scatter}
\end{figure}

\subsection{Real-world dataset 1: bike-sharing data}

In this section, we apply the proposed methods to the real-world dataset: trip data from a dock-based bike-sharing system.
We first introduce details of the dataset utilized in the experiment and then analyze the estimated real demand for the bike-sharing stations. 

\subsubsection{Data description}

We target the dock-based bike-sharing system in Manhattan, New York, USA. The trip data is provided by Citi Bike, the largest bike-sharing system in the USA. The open-source data includes the start and end stations of each trip, along with the corresponding times.
We focus on the real demand estimation during two peak periods from 8:00 to 9:00, and 16:00 to 17:00 on weekdays in March and April 2019. There are a total of 790 stations and 226530 trip records.

Firstly, for each bike-sharing station, the observed data $\mathcal{D} = \{(I_n, t^*_n, k_n)\}_{n=1}^N$ can be obtained. 
Similarly, for each time period, the observed drop-offs times $\boldsymbol{t_d}$ and total number of drop-offs $N_d$ are computed based on Equation \ref{eq:observed drop-offs}, while observed pick-up times $\boldsymbol{\tilde{t}_p}$ and total number of pick-ups $N_p$ are computed based on Equation \ref{eq:observed pick-ups}. 
We aim to recover the real demand at stations experiencing a shortage of bike supply. 

The experiments on synthetic datasets indicate that when there are no available vehicles at the initial time of a station, the censoring of either demand results in the final observed demand being close to each other. 
However, in real-world scenarios, it is possible for a station to have remaining vehicles from the previous time period or for rebalancing actions to have been performed during the current period. 
Therefore, when observed demand for pick-ups $N_p$ exceeds the demand for drop-offs $N_d$ in the data, it is likely attributable to these factors, highlighting an imbalance between supply and demand during that period. 
Conversely, if the $N_d$ significantly exceeds $N_p$, it suggests that the probability of real demand being censored during that period is minimal.
Consequently, we estimate the real demand for the stations satisfying the criterion $N_p / N_d \geq 0.8$ to exclude the cases that $N_d$ significantly exceeds $N_p$. 
The GVST observations are collected daily and then integrated into the final observations $\boldsymbol{y}=\{y_i\}^{N_g}_{i=1}$.

\begin{figure}[h]
    \centering
    \includegraphics[width=0.8\linewidth]{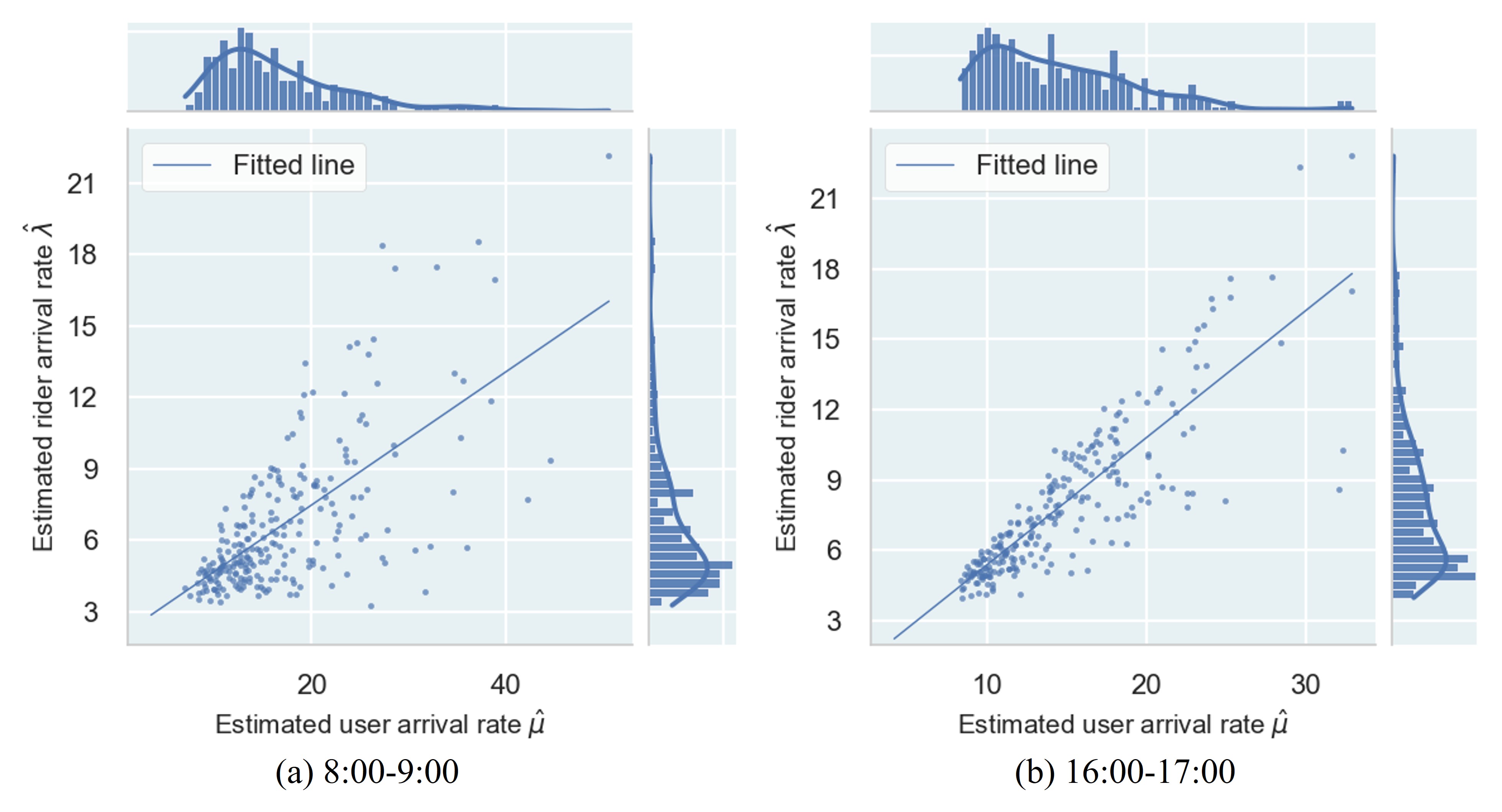}    \caption{Relationship between $\hat{\mu}$ and $\hat{\lambda}$.}
    \label{fig:BS linear}
\end{figure}

\begin{figure}[h]
    \centering    \includegraphics[width=0.8\linewidth]{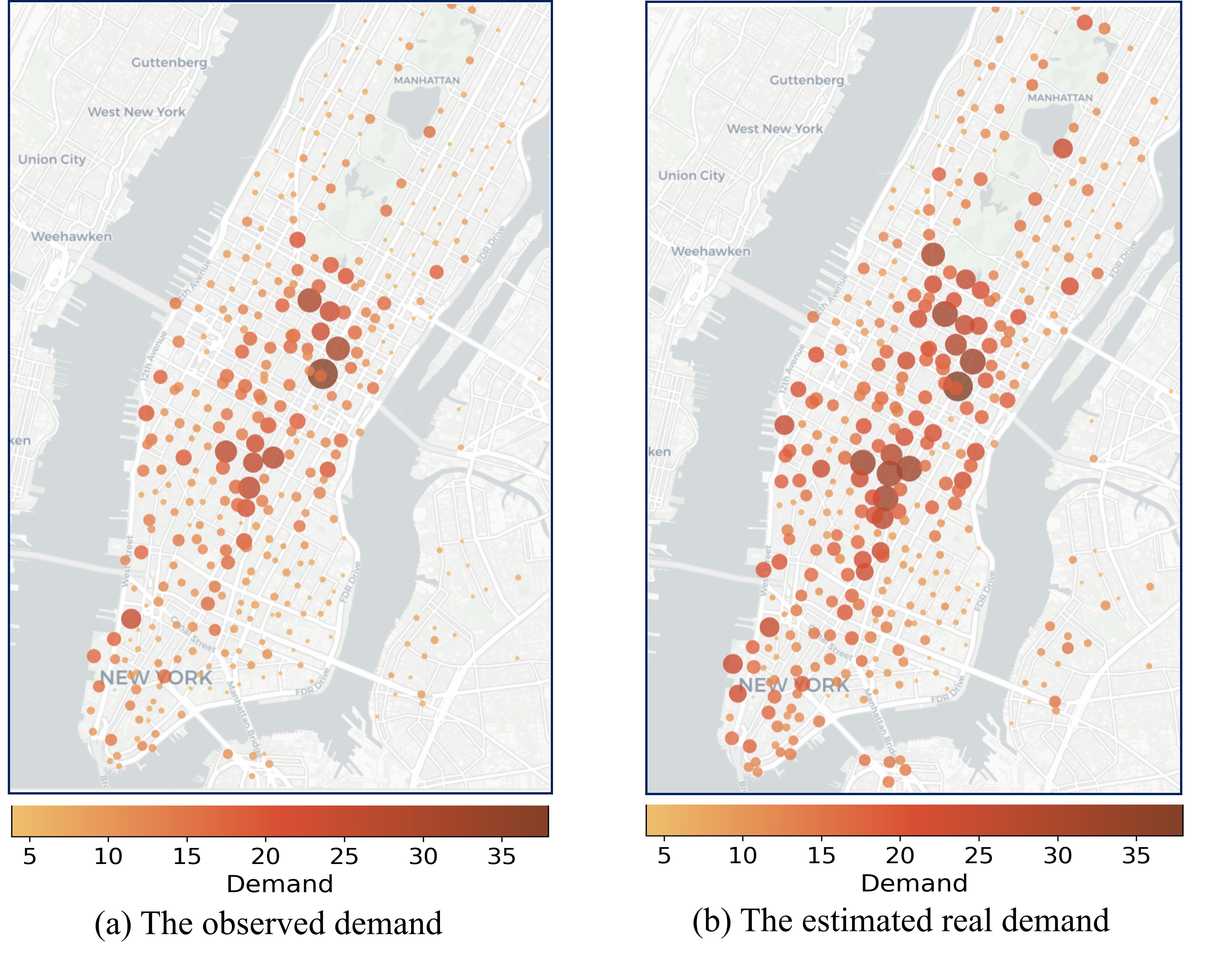}
    \caption{Comparison between observed demand and estimated real demand during evening peak. The circles represent the stations in the study area. The radius size and color intensity of the circles represent the demand for pick-ups in stations.}
    \label{fig:obs est}
\end{figure}

\subsubsection{Experimental results}

By applying the proposed estimation method, the estimated real demand for picking up bikes at each station can be obtained. 
The estimated results that pass the Kolmogorov-Smirnov (KS) test are treated as consistent and reliable in our study (KS test is discussed in Section \ref{sec:ks}).
Figure \ref{fig:bike scatter} (a) and \ref{fig:bike scatter} (b) illustrate the relationship between observed demand and real demand during the two peaks. Each point in the figure represents an individual station. 
The proximity of the data points to the diagonal line represents the degree of demand censoring at the stations. Points close to this line suggest that the observed demand closely matches the real demand. Conversely, points significantly deviating from this line highlight discrepancies between observed and real demand.
It can be seen that the observed demand at various stations is linearly related to the real demand for stations where the demand is censored.
Then we show the scatter plot of $\hat{\mu}$ versus $\hat{\lambda}$ for these stations in Figure \ref{fig:BS linear}.
The relationship between $\hat{\mu}$ and $\hat{\lambda}$ is predominantly linear during the evening peak. Specifically, for every increment in demand for pick-ups, the corresponding demand for drop-offs approximately increases by only 0.56 increments. 
This finding implies that the current system reaches a proportional balance. 
The linear relationship suggests a consistent degree of censoring, meaning that the proportion of the demand being censored is similar across stations of different sizes. 
It further suggests that the operator's distribution of stations and the setting of dock sizes are relatively reasonable. 
Additionally, it indirectly supports the validity and reasonableness of our estimation results.
The existence of the linear relationship can guide operators to allocate resources more precisely and scale the supply to match the unmet demand, thereby further optimizing operational performance.
However, the linear fitting result exhibits a poor fit during the morning peak. This could be attributed to the increased complexity of travel demand and rebalancing strategies during this period.

To further analyze the spatial correlations among the censoring of demand, we visualize the spatial distribution of the stockout rates for the stations in the Manhattan area.
Figures \ref{fig:obs est} (a) and (b) provide a preliminary comparison of the observed demand and real demand for bike-sharing stations during the two peaks.
The radius size and color intensity of the points represent the demand for pick-ups in stations. 
A larger radius indicates a higher demand volume, while a smaller radius indicates a lower demand. Similarly, darker colors correspond to higher demand, whereas lighter colors correspond to lower demand.

\begin{figure}[h]
    \centering    \includegraphics[width=0.8\linewidth]{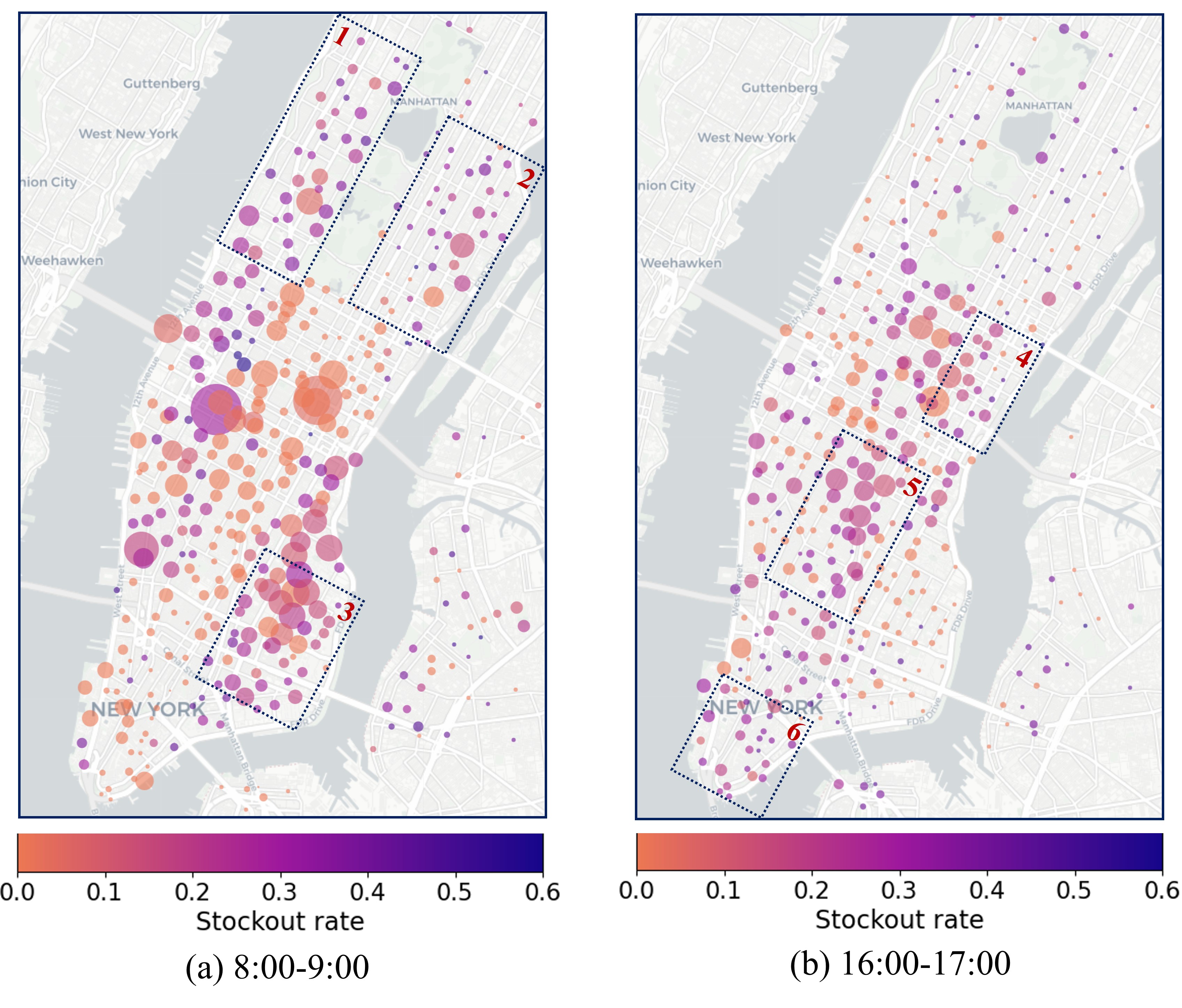}
    \caption{The stockout rates of bike-sharing stations in the NYC area. Different colors represent different stockout ratios. Each circle represents a station in the study area. The radius of each circle indicates the observed demand for pick-ups in the station.}
    \label{fig:NYC}
\end{figure}

In addition, the stockout ratio is introduced to illustrate the degree of censoring in the real demand, which is computed as $1-N_p / \mu$. 
It indicates the probability that a user arrives at a unit but leaves without picking up a vehicle \citep{xu2023locational}.
The distributions of obtained stockout rates for stations in the Manhattan area during the two peak periods are shown in Figure \ref{fig:NYC}.
Upon comparing the two time periods, it can be seen that the stockout rate distributions during the two peaks are approximately negatively correlated, especially for those stations facing large stockout rates.
Specifically, stations with higher stockout rates during the morning peak tend to exhibit lower stockout rates during the evening peak.
For the afternoon peak period shown in Figure \ref{fig:NYC} (b), the stockout rates are generally lower than those during the morning peak.
By examining the regions with high stockout rates during the morning peak, we found that they share the following characteristic: a higher concentration of residential points of interest (POIs).
Specifically, regions 1, 2, and 3 exhibiting generally higher stockout rates correspond to the Upper West Side, Upper East Side, East Village, and Lower East Side integration. These areas mainly consist of residential buildings such as high-rise apartments, tenement buildings, and new developments.
On the other hand, in the regions experiencing high stockout rates during the evening peak, a common characteristic observed is a higher density of commercial POIs.
As shown in Figure \ref{fig:NYC} (b), the stations in regions 4, 5, and 6 exhibit a concentrated higher stockout rate.
Region 4 is the Midtown East, which is known for its concentration of commercial buildings, while region 5 is the integration of the Soho area, Fifth and Sixth Avenue lined with numerous high-rise office buildings. Region 6 corresponds to the financial district with Wall Street and major financial institutions.
The above-mentioned results are consistent with the actual situations.
Since the users during the morning peak tend to pick up vehicles from the stations around residential regions, which leads to higher stockout rates. In contrast, users frequently pick up vehicles from stations near official and commercial areas, resulting in higher stockout rates in the morning peak hours.

\subsubsection{Comparison with closed-form solutions}
\begin{figure}[h]
    \centering    \includegraphics[width=0.8\linewidth]{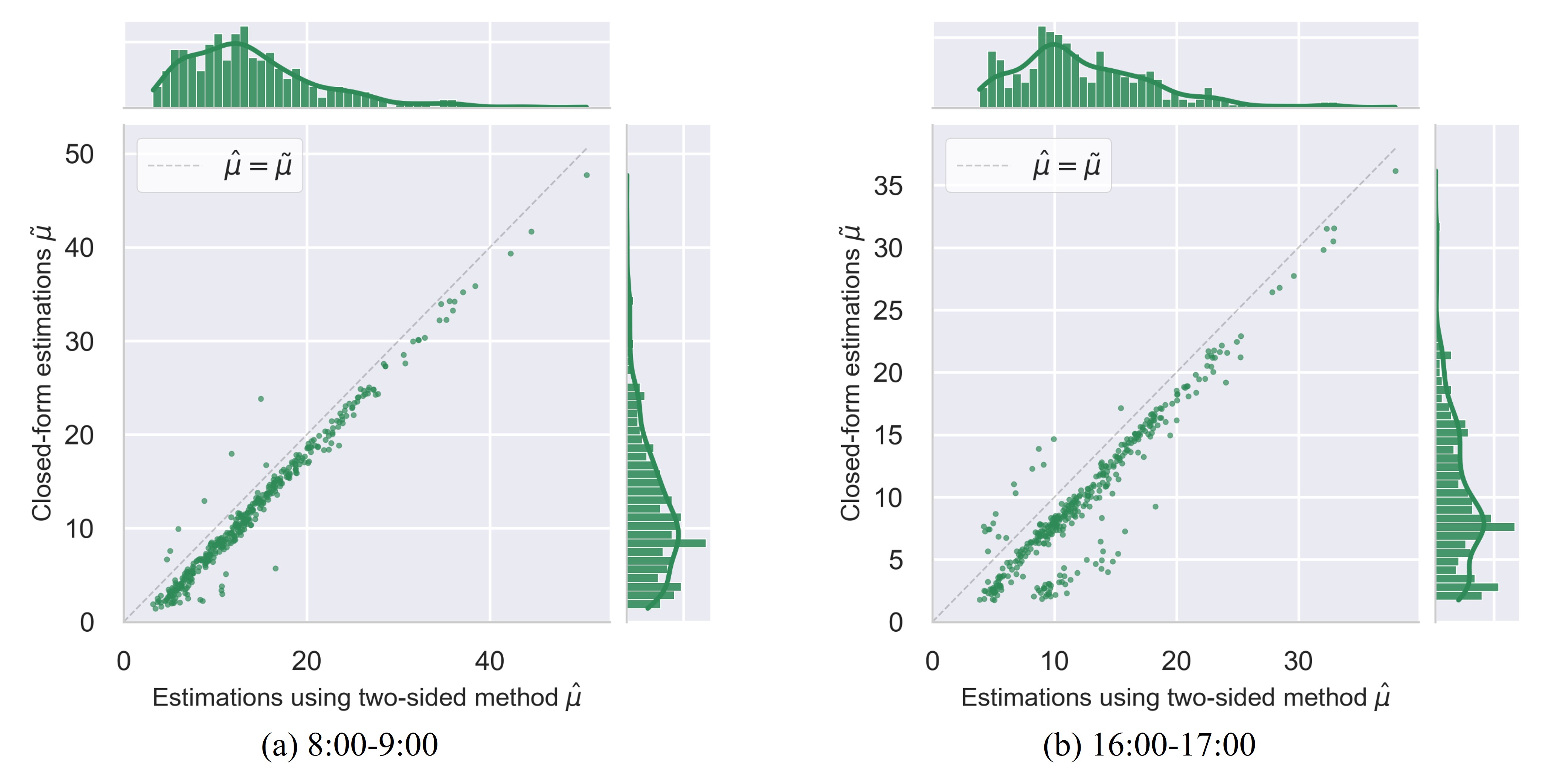}
    \caption{Relationship between two-sided and one-sided estimations (closed-form).}
    \label{fig:closed form}
\end{figure}

Next, we analyze the practicality of the closed-form solutions in the context of real-world datasets by comparing the results of the one-sided (closed-form) estimations with those obtained using the two-sided method. 
The two sets of estimation results are presented in Figure \ref{fig:closed form}, where the x-axis corresponds to the estimations derived from the two-sided method, and the y-axis represents the estimations derived from the one-sided estimations. 
It can be seen that most data points are located near the diagonal, which indicates that the one-sided estimations closely align with the two-sided estimation results.

Note that a significant number of estimations derived from the one-sided approach tend to slightly underestimate the real demand relative to the two-sided method. 
This observation is consistent with the experimental findings on synthetic datasets, which further validate the reliability of the one-sided estimations in approximating real-world demand patterns. 
Despite these occasional underestimations, the one-sided estimations remains a robust and efficient approximation, especially considering the simplicity and speed of its computational process.

This comparison highlights the practical value of the proposed one-sided method as a quick and reasonably accurate tool for estimating real demand. 
The ability to generate estimations that are computationally less demanding, yet still yield results within an acceptable margin of error, makes the one-sided method an appealing choice for practitioners in real-time decision-making scenarios.
Moreover, the one-sided method can serve as a useful rule of thumb for managers, facilitating faster demand estimation without the need for more resource-intensive methods. 
Although the two-sided method may provide slightly more accurate results, the one-sided approach offers a practical balance between computational efficiency and estimation accuracy. 
In scenarios where rapid decision-making is crucial and exact demand estimations are not strictly necessary, the one-sided solution can be a valuable tool in managing micro-mobility systems effectively.

\subsection{Real-world dataset 2: shared e-scooter data}

In this section, we extend the proposed methods to dockless systems by applying real-world data from an e-scooter sharing service.
In addition, sensitivity analysis regarding the grid size within the e-scooter system is also conducted to justify the robustness of the proposed method.

\subsubsection{Data description}

The dockless shared e-scooter system in Berlin, Germany, one of the top cities in the e-scooter market is studied in this section.
The e-scooter data in Berlin was sourced through the public Application Programming Interface (API) provided by the operator Voi Technology. 
The data in April 2021 was queried from the API. 
The raw data captures the locations and times of e-scooters when they are idle. The trip data for each e-scooter can be derived by linking the end of one idle period to the start of the next.
After data processing, each record comprises information including e-scooter ID, start and end location (longitude and latitude), and start and end timestamp. 
We focus on the demand estimation during the evening peak, from 17:00 to 18:00 on weekdays.
First, we divide the urban area of Berlin into equal-sized grids (800m $\times$ 800m). The estimation results with different grid sizes are discussed in section \ref{sec: SA}.
By geographically matching the locations in the records and the grids, the trip data can be transformed into forms containing the starting and ending grid along with the times.
Similar to the dock-based bike-sharing systems, the observed demand for pick-ups and drop-offs within each grid can be calculated. Subsequently, the final GVST observations are computed based on Equation \ref{eq:gvst}. 
The capacity of each grid is defined as the maximum number of available vehicles within the grid in the historical data.

\begin{figure}[h]
    \centering    \includegraphics[width=0.8\linewidth]{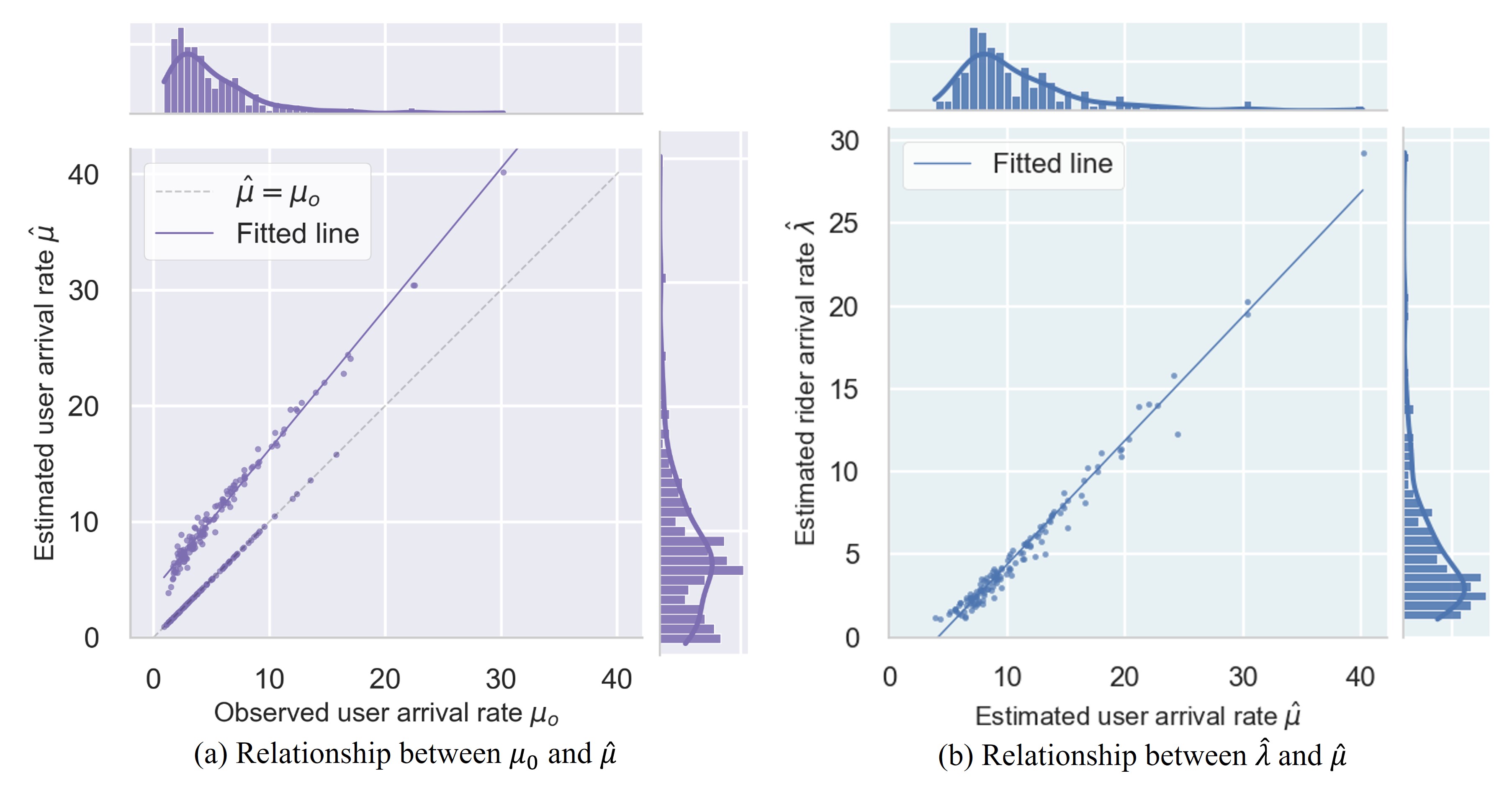}
    \caption{The fitted linear lines during the evening peak.}
    \label{fig:scatter}
\end{figure}

\subsubsection{Experimental results}

Figure \ref{fig:scatter} (a) presents the estimated real demand for the grids during the evening peak. 
A linear regression analysis is performed to model the relationship between the $\hat{\mu}$ and $N_p$.
The fitted line suggests that the degree of censorship tends to increase as the observed demand decreases. 
In addition, we also conduct a linear regression analysis to describe the relationship between $\hat{\lambda}$ and $\hat{\mu}$. The fitted result in Figure \ref{fig:scatter} (b) shows a significant linear relationship between the two variables. 
For every unit of increment in demand for pick-ups, the corresponding demand for drop-offs increases by approximately 0.75. This finding indicates that the system achieves a partial supply-demand equilibrium, and the proportion of the demand being censored remains consistent across different grids. Additionally, this implies that the distribution of shared fleets of e-scooters and rebalancing strategies from the operator is reasonable but can be further improved to meet the latent real demand.

\begin{figure}[h]
    \centering    \includegraphics[width=0.9\linewidth]{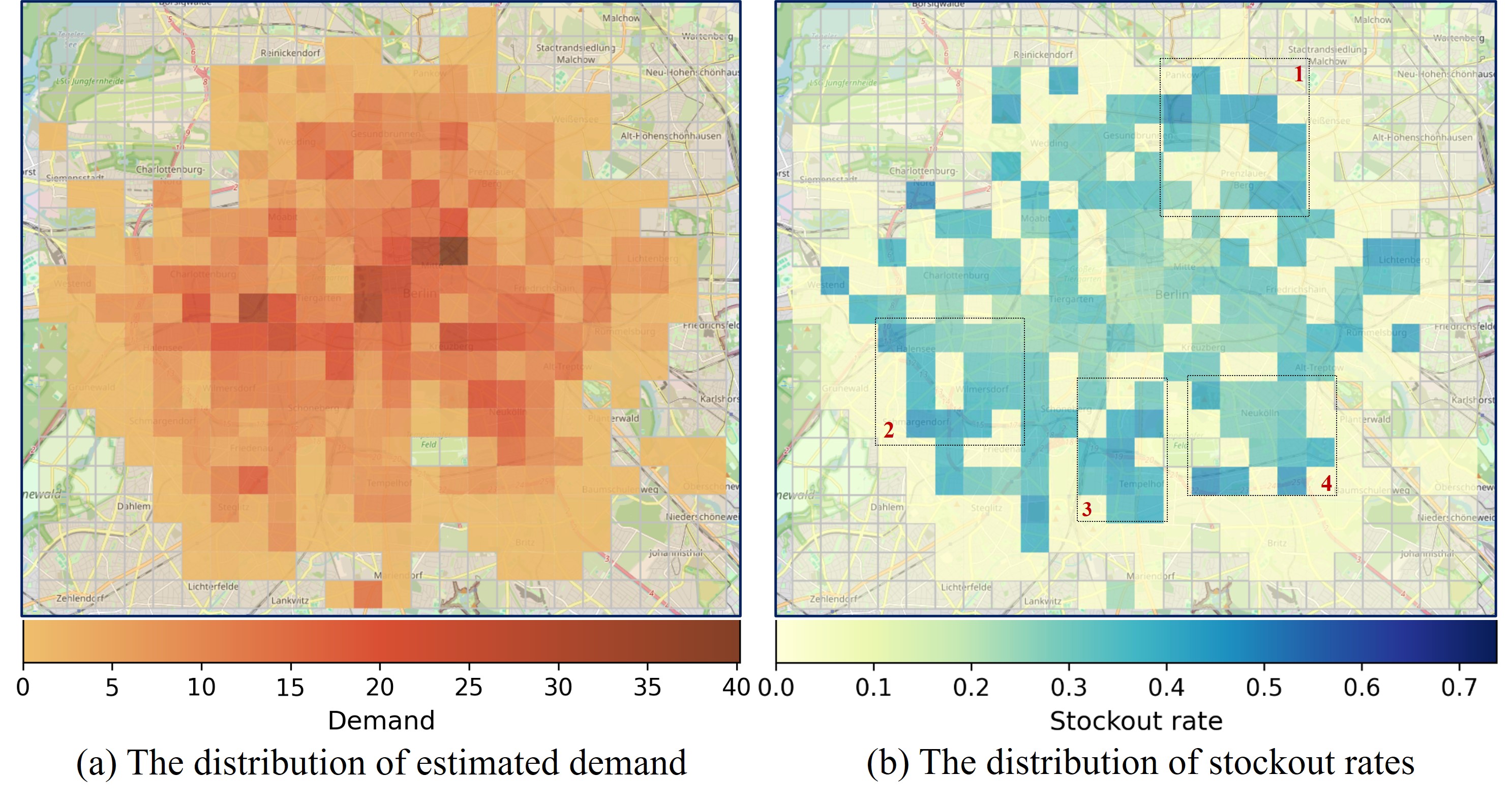}
    \caption{The estimated demand and stockout rates during the evening peak.}
    \label{fig:Berlin}
\end{figure}

We then visualize the spatial distribution of the estimated demand and the corresponding stockout rates in the Berlin area during the evening peak in Figure \ref{fig:Berlin}. For the estimated demand shown in Figure \ref{fig:Berlin} (a), the deeper color means a larger estimated demand $\hat{\mu}$ for shared e-scooter services within the grid.
For the distribution of stockout rates in Figure \ref{fig:Berlin} (b), the deeper color means a larger stockout rate and a larger proportion of demand is unmet within the grid.
The results suggest that supply-demand imbalances are more frequent in semi-popular grids, which are generally away from the city center during the evening peak. 
This may be attributed to the lack of rebalancing strategies for suburbs.
By examining the land use of the grids with higher stockout rates, the grids containing more commercial and working areas are under more severe censoring during the evening peak.
The reason may be that the concentrated departure of people from these grids creates a high demand, while residential zones may experience an accumulation of e-scooters.
For example, region 1 includes several work zones and commercial districts (e.g., Prenzlauer Berg and Gesundbrunnen) in Pankow, a borough in Berlin. Region 2 contains various offices and business stores in Halensee, Fehrbelliner Platz surrounded by a notable work zone and commercial streets in Schmargendorf.
Regions 3 and 4 correspond to the Körnerkiez in the North, which is known for its rich cultural and commercial activities and Tempelhofer Damm, a central business hub. The interface between the two regions is also rapidly developing many emerging enterprises.

\subsubsection{Sensitivity analysis regarding the grid size}
\label{sec: SA}

In this section, we perform the sensitivity analysis to examine the effect of grid size on the estimation result of real demand.
Smaller grids may provide more detailed and localized demand patterns, while larger grids offer a broader overview. 
Performing the sensitivity analysis of grid size enhances the generalizability of the demand estimation models by establishing a robust foundation for extending the estimation method to different regions and conditions.
To this end, we consider the equal-sized grids of lengths ranging from 500 meters (m) to 1000m, increasing in increments of 100m to investigate how the grid size impacts the estimation outcome. 
Different grid sizes affect the data generation for the observed data $\mathcal{D} = \{(I_n, t^*_n, k_n)\}_{n=1}^N$, which further influences the observations of GVST $\boldsymbol{y}=\{y_i\}^{N_g}_{i=1}$ and ultimately the estimated results.

\begin{figure}[h]
    \centering    \includegraphics[width=0.95\linewidth]{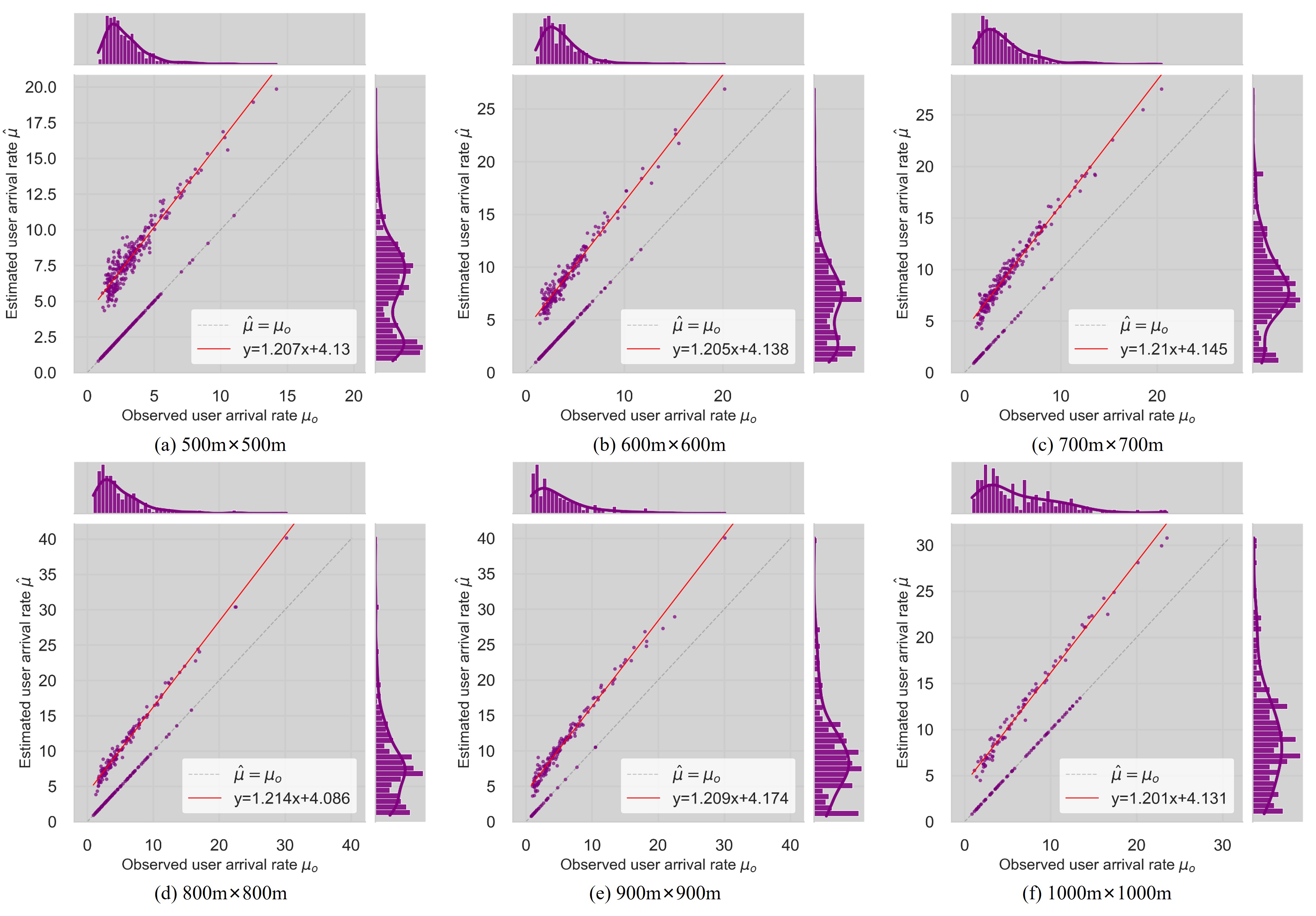}
    \caption{Sensitivity analysis.}
    \label{fig:SA}
\end{figure}

Then we compare the estimated results from different grid sizes, and the results are presented in Figure \ref{fig:SA}. The linear regression is also conducted on the estimated demand for grids. It can be seen that the slope of the fitted lines remains consistent across different grid sizes, indicating the robustness of our method. 
Therefore, the stability regarding the grid size suggests that the grid scale does not significantly affect the estimation result of the real demand.

\subsection{Goodness of fitting test}
\label{sec:ks}

In this section, we assess the compatibility between the observed GVST from the two real-world datasets and the proposed GVST distribution parameterized by the estimated $\hat{\mu}$. To evaluate the accuracy of the estimated model in capturing the behavior of the GVST observations, we employ the Kolmogorov-Smirnov (KS) test \citep{massey1951kolmogorov}.
The KS test is a non-parametric method that compares an empirical cumulative distribution function with a specified theoretical cumulative distribution function.
The null hypothesis of the KS test assumes that the GVST observations are drawn from the distribution $f_Y(y|\hat{\lambda}, \hat{\mu}, K)$. 
The null hypothesis is rejected at the 5\% significance level when the associated $p$-value is less than $0.05$.
The results show 83.7\% of the stations in the bike-sharing data and 78.1\% of the zones in the shared e-scooter data fail to reject the null hypothesis, thereby indicating that the GVST observations are consistent with the estimated distribution. 
We found that, for the cases where the null hypothesis is rejected, more than 60 \% in the bike-sharing data and 50 \% in the shared e-scooter data involve fewer than 100 GVST observations. 
Therefore, the insufficient number of GVST observations appears to be the primary reason for the lack of fit.
For other cases, the observed poor fit may be attributed to the inconsistent demand across different days and the effects of rebalancing strategies. 
These factors may disrupt the assumptions of Poisson arrival processes and lead to biased estimations.

\section{Conclusion}
\label{sec:Conclusion}

This study addresses the challenge of real demand estimation for shared micro-mobility services, which is often hindered by the spatial-temporal imbalance between supply and demand. 
We propose a novel analytic approach by introducing an FQM, which flips the traditional roles of supply and demand to better capture the dynamics of these services. Central to our approach is the introduction of the GVST as a random variable observable from trip records, allowing us to transform the real demand estimation problem into an inverse queueing problem.
The proposed two-sided and one-sided estimation methods diverge from existing literature by avoiding machine learning or optimization-based techniques and instead relying on analytical solutions to estimate real demand.
Our experimental results, based on both synthetic datasets and real-world data from bike-sharing and shared e-scooter systems, demonstrate the effectiveness of these methods in accurately recovering real demand.
State-of-the-art models are compared to further validate the effectiveness of our methods. 
The results demonstrate that both methods outperform other baseline models, which indicates their great potential in recovering real demand for shared micro-mobility services.
Specifically, the two-sided method provides more reliable estimations compared to the one-sided method.
Importantly, while the one-sided approach yields a slightly less precise result, its closed-form solution remains sufficiently accurate, providing a practical and reliable tool for managers and practitioners in the field.
Extensive experiments with real-world datasets from a dock-based bike-sharing system and a dockless e-scooter system demonstrate that the estimation results align well with the actual traffic situations. In the evening peak hours, the estimated real demand of units suggests a significant linear relationship for both bike-sharing and e-scooter systems. It indicates that a partial supply-demand equilibrium is achieved by the current operational strategies of the two operators.
Overall, this research significantly contributes to the field of micro-mobility service management by offering an innovative, analytically grounded method for real demand estimation. 
The proposed methods offer an effective alternative to traditional machine learning or optimization-based techniques, with the one-sided closed-form solution providing a simple yet reliable approximation for real demand. 
This approach not only enhances the accuracy of demand estimation but also offers a valuable tool for improving decision-making processes in micro-mobility planning and operations.

For future research directions, it is highly practical and significant to incorporate the temporal and spatial attributes when estimating the real demand for micro-mobility services. 
The demand during a given time period is supposed to be related to the previous time periods due to the influence of historical trends and patterns.
Also, we aim to further relax the assumption that the arrival rate is constant during a period.
For instance, the time-varying Poisson processes can be introduced to relax the assumption of constant arrival rates. However, how to investigate the distribution of GVST under such a scenario is worth further discussion and analysis.
Modeling the user and vehicle arrival process with rates changing over time is more consistent with real-world cases. 
Integrating the above time-varying arriving process and the temporal dependencies into the proposed FQM is challenging yet valuable.
The distribution of real demand is related to the spatial correlations, and the lost demand of one unit may be transformed to adjacent units. In future works, we will examine how to integrate such spatial correlations and the transition processes between units into the FQM. 
Furthermore, it would be helpful to include the effects brought by other traffic modes by integrating the multi-source data. For instance, the unmet demand from micro-mobility may be tractable by analyzing the ride-hailing demand simultaneously.

\bibliography{main}

\appendix
\section{The case when the $\lambda$ and $\mu$ are equal}
\label{App: A}

If $\lambda$ is equal to $\mu$, the steady-state probability of the inventory in the system is given as follows:

\begin{equation}
\label{eq:pn}
    P_x=\frac{1}{K+1}.
\end{equation}

Then the CDF of GVST when $\lambda$ equals to $\mu$ can be derived as:

\begin{equation}
\begin{aligned} 
\label{eq:cdf app}
F_Y(y) = 1-\sum_{x=0}^{K-1}\left(\sum_{i=0}^x\frac{(\mu y)^i}{i!}e^{-\mu y}\right)\frac{P_x}{1-P_K} 
= 1-\sum_{x=0}^{K-1}\left(\sum_{i=0}^x\frac{(\mu y)^i}{i!}e^{-\mu y}\right)\frac{1}{K}.
\end{aligned} 
\end{equation}

Subsequently, the PDF of GVST $f_Y(y)$ can be calculated as

\begin{equation}
\label{eq:pdf app}
f_Y(y)= \sum_{x=0}^{K-1}\mu\frac{(\mu y)^x}{x!}e^{-\mu y}\frac{P_x}{1-P_K} = \sum_{x=0}^{K-1}\mu\frac{(\mu y)^x}{x!}e^{-\mu y}\frac{1}{K},
\end{equation}

\section{Proofs of Proposition \ref{prop:two}}
\label{App: multi-solutions}

We begin by examining the diagonal elements of the Hessian matrix, specifically $\frac{\partial^2 \mathcal{L}}{\partial \mu^2}$ and $\frac{\partial^2 \mathcal{L}}{\partial \lambda^2}$ are formulated as:

\begin{equation}
\frac{\partial^2 \mathcal{L}}{\partial \mu^2}= 
- \frac{N_g}{(\mu - \lambda)^2}
- \frac{N_gK \lambda^K (\lambda^K - (1+K)\mu^K)}{\mu^2 (\mu^K-\lambda^K)^2}
\label{eq:partial2 mu}
\end{equation}

\begin{equation}
\frac{\partial^2 \mathcal{L}}{\partial \lambda^2}
= \frac{-N_g}{(\mu-\lambda)^2} 
+ \frac{N_gK\lambda^{K-2}(K\mu^K-\mu^K+\lambda^K)}{(\mu^K-\lambda^K)^2} 
+ \sum_{i=1}^{N_g}\left(\frac{\sum_{k=2}^{K-1}\frac{y_i^k \lambda^{k-2}}{(k-2)!}}{\sum_{k=0}^{K-1}\frac{(\lambda y_i)^k}{k!}} 
- \left(\frac{\sum_{k=1}^{K-1}\frac{y_i^k\lambda^{k-1}}{(k-1)!}}{\sum_{k=0}^{K-1}\frac{(\lambda y_i)^k}{k!}}\right)^2\right)
\label{eq:partial2 lambda}
\end{equation}

Next, we consider the case when $\lambda=100$, $\mu=180$, $K=20$ with the corresponding GVST observations $\boldsymbol{y}=\{y_i\}^{5000}_{i=1}$, By substituting various parameter values for $\lambda$ and $\mu$ into Equations \ref{eq:partial lambda} and \ref{eq:partial mu}, we obtain the following result: when $\lambda=100$ and $\mu=300$, the value of $\frac{\partial^2 \mathcal{L}}{\partial \lambda^2}$ is -0.1249.
This outcome contradicts the fundamental property of negative semidefinite matrices, which dictates that the diagonal elements must be non-positive.
Consequently, the Hessian matrix $H_t$ cannot be classified as negative semidefinite. This suggests that the likelihood function may exhibit multiple local maxima and the global maximum may not be identifiable.

\section{System of equations}
\label{App:system of equation}
By setting the partial derivatives of $l(\lambda, \mu | K,\boldsymbol{y})$ with respect to $\lambda$ and $\mu$ to 0, we can obtain the following system of equations:

\begin{equation}
\begin{aligned} 
\label{eq:system of equation}
- \frac{N_g}{\mu-\lambda} + \frac{N_gK\lambda^{K-1}}{\mu^K-\lambda^K} 
+ \sum_{i=1}^{N_g}\frac{\sum_{k=0}^{K-1}\frac{y_i^k\lambda^{k-1}k}{k!}}{\sum_{k=0}^{K-1}\frac{(\lambda y_i)^k}{k!}}
& = 0 \\
- \sum_{i=1}^{N_g} y_i + \frac{N_g}{\mu-\lambda} - \frac{NK\mu^{K-1}}{\mu^K-\lambda^K} + \frac{N_g K}{\mu} & = 0
\end{aligned} 
\end{equation}

\end{document}